\documentclass{article}

\usepackage[english]{babel}

\usepackage[letterpaper,top=2cm,bottom=2cm,left=2.5cm,right=2.5cm,marginparwidth=1.75cm]{geometry}

\usepackage{amsmath}
\usepackage{amssymb}
\DeclareMathOperator*{\argmax}{arg\,max}
\usepackage{graphicx}
\usepackage[colorlinks=true, allcolors=black]{hyperref}
\usepackage{bm}
\usepackage{float}
\usepackage{hyperref}
\usepackage{bbold}
\usepackage{verbatim}
\usepackage[table]{xcolor}
\definecolor{highlight}{RGB}{135,206,250} 
\usepackage{subfigure}
\usepackage{caption}
\usepackage{multirow}
\usepackage{booktabs}
\usepackage{natbib}
\usepackage[flushleft]{threeparttable}
\usepackage{pdflscape}
\usepackage{everypage}
\usepackage{textcomp}
\usepackage{tikz}    
\newcommand{\hide}[1]{}
\usetikzlibrary{shapes.geometric, arrows, shadows, fadings}
\tikzstyle{startstop} = [rectangle, rounded corners, minimum width=3.5cm, minimum height=1.2cm, text centered, draw=black, fill=gray!20]
\tikzstyle{arrow} = [thick,->,>=stealth]
\usepackage{setspace}

\newcommand{\Lpagenumber}{\ifdim\textwidth=\linewidth\else\bgroup
  \dimendef\margin=0 
  \ifodd\value{page}\margin=\oddsidemargin
  \else\margin=\evensidemargin
  \fi
  \raisebox{\dimexpr -\topmargin-\headheight-\headsep-0.5\linewidth}[0pt][0pt]{%
    \rlap{\hspace{\dimexpr \margin+\textheight+\footskip}%
    \llap{\rotatebox{90}{\thepage}}}}%
\egroup\fi}
\AddEverypageHook{\Lpagenumber}%

\title{\Large \bf A Bayesian joint model of multiple nonlinear longitudinal and competing risks outcomes for dynamic prediction in multiple myeloma: joint estimation and corrected two-stage approaches \bigskip}

\author{\normalsize\textbf{Danilo Alvares}$^{1,*}$, \textbf{Jessica K. Barrett}$^{\bm 1}$,
\textbf{Fran\c{c}ois Mercier}$^{2}$, \textbf{Spyros Roumpanis}$^{2}$, \smallskip \\ 
\normalsize\textbf{Sean Yiu}$^{3}$, \textbf{Felipe Castro}$^{\bm 2}$, \textbf{Jochen Schulze}$^{\bm 2}$, \textbf{and} \textbf{Yajing Zhu}$^{\bm 2,**}$ \bigskip \\ \normalsize
$^{1}$MRC Biostatistics Unit, University of Cambridge, U.K. \\ \normalsize
$^{2}$F. Hoffmann-La Roche Ltd, Basel, Switzerland \\ \normalsize
$^{3}$Roche Products Ltd, Welwyn Garden City, U.K. \bigskip \\ \normalsize
$^{*}${\it email:} danilo.alvares@mrc-bsu.cam.ac.uk \\ \normalsize
$^{**}${\it email:} yajing.zhu@roche.com
}

\date{}

\begin{document}

\maketitle

\begin{abstract}
\noindent Predicting cancer-associated clinical events is challenging in oncology. In Multiple Myeloma (MM), a cancer of plasma cells, disease progression is determined by changes in biomarkers, such as serum concentration of the paraprotein secreted by plasma cells (M-protein). Therefore, the time-dependent behaviour of M-protein and the transition across lines of therapy (LoT) that may be a consequence of disease progression should be accounted for in statistical models to predict relevant clinical outcomes. Furthermore, it is important to understand the contribution of the patterns of longitudinal biomarkers, upon each LoT initiation, to time-to-death or time-to-next-LoT. Motivated by these challenges, we propose a Bayesian joint model for trajectories of multiple longitudinal biomarkers, such as M-protein, and the competing risks of death and transition to next LoT. Additionally, we explore two estimation approaches for our joint model: simultaneous estimation of all parameters (joint estimation) and sequential estimation of parameters using a corrected two-stage strategy aiming to reduce computational time. Our proposed model and estimation methods are applied to a retrospective cohort study from a real-world database of patients diagnosed with MM in the US from January 2015 to February 2022. We split the data into training and test sets in order to validate the joint model using both estimation approaches and make dynamic predictions of times until clinical events of interest, informed by longitudinally measured biomarkers and baseline variables available up to the time of prediction. \medskip 
\end{abstract}

\section{Introduction} \label{sec:intro}

Recent trends in personalised healthcare have motivated great interest in the individual dynamic risk prediction of survival and other clinically important events by using baseline characteristics and the course of disease progression \citep{barrett2017, ferrer2019, ren2021, parr2022}. In particular, studies of Multiple Myeloma (MM, the second most common hematological cancer) have identified several risk factors that may help to predict the disease course \citep{abdallah2023, zhang2023}. In this type of blood cancer, malignant plasma cells accumulate in the bone marrow and secrete a monoclonal protein/paraprotein (also known as M-protein). The time-dependent assessment of M-protein concentration through serum protein electrophoresis (SPEP/M-spike) and/or involved free light chains (FLC), e.g. through FreeLite$^\text{\scriptsize\sffamily\textregistered}$ test, may provide useful information to the treating physician about the individual risk of a patient to experience either one of two clinical events of interest: death or start of a new line of therapy (LoT) \citep{kumar2017}.

To understand the dynamic interplay between longitudinal biomarkers and their associations with clinical outcomes, we propose a new Bayesian joint model that appropriately accommodates different characteristics of MM data. Specifically, upon each LoT initiation, the temporal profiles of M-spike and FLC are nonlinear, and could adequately be characterised by a bi-exponential model \citep{stein2008,claret2009}. This model presents three components (baseline, growth rate, decay rate parameters) to summarise the longitudinal trajectory and to explain the time until clinical events of interest (death or start of next LoT). These two clinical events are modelled as competing risks, in which we use a proportional cause-specific hazard specification \citep{putter2020}.

The simultaneous estimation of all parameters in a joint model is computationally intensive due to the complexity of approximating posterior distributions from multiple nonlinear longitudinal submodels sharing information with a competing risks submodel \citep{hickey2016, mauff2020}. As an alternative approach, we explore the corrected two-stage approach proposed by \citet{alvares2023}. This approach reduces computational complexity by estimating the submodels separately and produces results similar to those of simultaneous estimation due to a bias correction mechanism incorporated in the second stage. Previous works have shown such inferential similarity between both approaches \citep{alvares2023, alvares2024a}, but it is unclear whether they also produce similar predictions. Thus, we also intend to shed light on this topic through comparisons using predictive metrics and individual dynamic predictions. Hence, a corrected two-stage proposal is compared to the joint estimation approach using a retrospective cohort study of patients diagnosed with MM and who received at least one LoT between January 2015 and February 2022. 

The rest of the work is organised as follows. Section~\ref{sec:data} describes a multiple myeloma retrospective cohort study from the US nationwide Flatiron Health database. Driven by such data, Section~\ref{sec:model} presents a Bayesian joint model of multiple nonlinear longitudinal and competing risks outcomes. Section~\ref{sec:approaches} introduces the joint estimation and corrected two-stage approaches, model performance evaluation criteria, and a dynamic risk prediction scheme. Section~\ref{sec:results} compares both estimation approaches applied to multiple myeloma data. The work ends with a discussion in Section~\ref{sec:discuss}.

\section{Multiple Myeloma Data} \label{sec:data}

Personalised patient management in diseases like MM is currently challenging due to disease heterogeneity and shortcomings of existing models to accurately predict patients at high risk of clinical events of interest, e.g. early relapse in MM \citep{vandevelde2007, lahuerta2008, martinez-lopez2011, rees2024}. Motivated by this context, we leveraged de-identified patient-level data from Flatiron Health electronic health record (EHR)-derived database of patients diagnosed with MM in the US. The Flatiron Health database is a nationwide longitudinal, demographically, and geographically diverse database derived from EHR data \citep{ma2023}. In totality, it includes de-identified data from over 280 cancer clinics (approximately 800 sites of care), representing more than 2.4 million patients with active cancer in the US \citep{kumar2021}. The majority of patients in the database originate from community oncology settings; relative community/academic proportions may vary depending on study cohort. The patient-level data in EHRs includes structured data (e.g. laboratory values and prescribed drugs) in addition to unstructured data collected via technology-enabled chart abstraction from physician's notes and other documents (e.g. biomarker reports and discharge summaries) \citep{birnbaum2020}. 

For our analyses, the follow-up period is defined from 1st January 2015 to 28th February 2022. Patients included in the database were diagnosed with MM on or after 1st January 2015 and presented at least two visits in the Flatiron Health system. Other eligibility criteria are: (i) at least 18 years of age at MM diagnosis, (ii) no longer than 60 days between initial diagnosis and first activity (visit or LoT initiation), (iii) more than three months on-treatment before the end of study follow-up, (iv) and no malignancies before MM diagnosis. Thus, 5490 patients formed the sample of newly diagnosed MM patients who met all eligibility criteria. Table~\ref{table:transitions} shows the number of patients who died, transitioned to next LoT, or were censored (neither experienced death nor started next LoT) and their respective median time until the event within each LoT (LoT in the database was oncologist-defined, rule-based). For example, of the 5490 patients who started LoT 1, 843 (15\%) died during LoT 1, 2775 (51\%) changed to LoT 2, 1872 (34\%) were censored during LoT 1, and their respective median time on LoT 1 was 210, 266, and 198 days. Note that we grouped LoT 5 and beyond, into LoT 4 due to the small number of patients (e.g. 362 in LoT 5) and to avoid issues in model parameters estimation. Clinically, such grouping is also acceptable as the combination of treatments in these later lines are in general less effective as a whole. In practice, this means that from LoT 4 onwards, only death or censoring can occur. Table~\ref{table:transitions} also shows a summary of the distribution of the number of M-spike and FLC measurements per patient by LoT.

\begin{table}[htp!]
\caption{Number of cases, percentages, and median times (in days) of patients who died, transitioned to next line of therapy (LoT), and were censored within each LoT $l=1,2,3,4$, as well as the summary of the distribution of the number of M-spike and FLC measurements per patient by LoT.}
\label{table:transitions}
\centering
\begin{threeparttable}
\begin{tabular}{cccccc}
\hline
{\bf Status}             & {\bf Summary} & {\bf LoT $\bm{l=1}$} & {\bf LoT $\bm{l=2}$} & {\bf LoT $\bm{l=3}$} & {\bf LoT $\bm{l=4}$} \\
\hline
\multirow{2}{*}{Death}   & Cases (\%)   & 843 (15) & 400 (14) & 246 (18) & 290 (41) \\
                         & Median       & 210      & 189      & 127      & 268      \\
\hline
\multirow{2}{*}{LoT $l+1$} & Cases (\%) & 2775 (51) & 1401 (51) & 716 (51) &  \\
                           & Median     & 266       & 226       & 174      &  \\
\hline
\multirow{2}{*}{Censored} & Cases (\%)  & 1872 (34) & 974 (35) & 439 (31) & 426 (59) \\
                          & Median      & 198      & 292      & 290      & 455      \\
\hline
\multicolumn{2}{c}{Total number of patients$^\ddagger$}      & 5490     & 2775     & 1401     & 716      \\
\hline
{\bf Biomarker}          & {\bf Summary} & {\bf LoT $\bm{l=1}$} & {\bf LoT $\bm{l=2}$} & {\bf LoT $\bm{l=3}$} & {\bf LoT $\bm{l=4}$} \\
\hline
\multirow{5}{*}{M-spike} & Min           & 1     & 1     & 1     & 1     \\
                         & Quartile 1    & 2     & 2     & 2     & 2     \\
                         & Median        & 4     & 4     & 3     & 3     \\
                         & Quartile 3    & 6     & 7     & 7     & 7     \\
                         & Max           & 97    & 53    & 66    & 43    \\
\hline
\multicolumn{2}{c}{Total number of patients$^\dagger$}      & 3673     & 1707     & 867     & 438      \\
\hline
\multirow{5}{*}{FLC}     & Min           & 1     & 1     & 1     & 1     \\
                         & Quartile 1    & 2     & 2     & 2     & 2     \\
                         & Median        & 4     & 5     & 4     & 4     \\
                         & Quartile 3    & 8     & 10    & 9     & 8     \\
                         & Max           & 78    & 76    & 69    & 55    \\
\hline
\multicolumn{2}{c}{Total number of patients$^\dagger$}      & 4053     & 1947     & 982     & 484      \\
\hline
\end{tabular}
  \begin{tablenotes}
    \item[*]The difference between $\ddagger$ and $\dagger$ indicates the number of patients with no record of either M-spike or FLC.
  \end{tablenotes}
\end{threeparttable}
\end{table}

In addition to times until clinical events of interest, the Flatiron Health database also provides observations for the longitudinal biomarkers that predict the probability of the clinical event of interest occurring. Here we use SPEP/M-spike results to quantify the concentration of M-protein in serum (g/L) and FLC to quantify the concentration of involved light chains (either kappa or lambda, g/L). Due to varying clinical guidelines and practices, a large number of patients may not always have both of these biomarkers or even different recording frequencies. Hence, for patients with both kappa-FLC and lambda-FLC, the chain of the higher initial value is followed through (see examples in Web Figure~\ref{fig0}).

Baseline characteristics collected at initial diagnosis are available for the categorical variables: sex, ethnicity, Eastern Cooperative Oncology Group (ECOG), and International Staging System (ISS) (see a descriptive summary in Web Table~\ref{table:catdescript}), as well as for the continuous variables: age, albumin  (serum, g/L), beta-2-microglobulin (B2M, serum, mg/L), creatinine (serum, mg/dL), hemoglobin (g/dL), lactate dehydrogenase (LDH, serum, U/L), lymphocyte (count, $\times 10^{9}$/L), neutrophil (count, $\times 10^{9}$/L), platelet (count, $\times 10^{9}$/L), immunoglobulin A (IgA, serum, g/L), IgG (serum, g/L), IgM (serum, g/L) (see a descriptive summary in Web Table~\ref{table:contdescript}). From LoT 2 onwards, we can use the time spent in the previous LoT as an explanatory variable. Except for age and time spent in the previous LoT, all continuous variables have missing data. To handle such variables, we apply a log transformation to reduce asymmetry, a standardisation so that their scales are similar, and a simple imputation to fill in missing observations (see Web Table~\ref{table:contdescript}).

For each LoT, we randomly select 80\% of the data to be the training set (used to fit, calibrate, and validate the model) and the remaining 20\% comprise the test set (a hold-out dataset to evaluate the model performance and make predictions). Web Figure~\ref{fig01} illustrates our train-test split strategy by LoT.

\section{The Bayesian Joint Model} \label{sec:model}

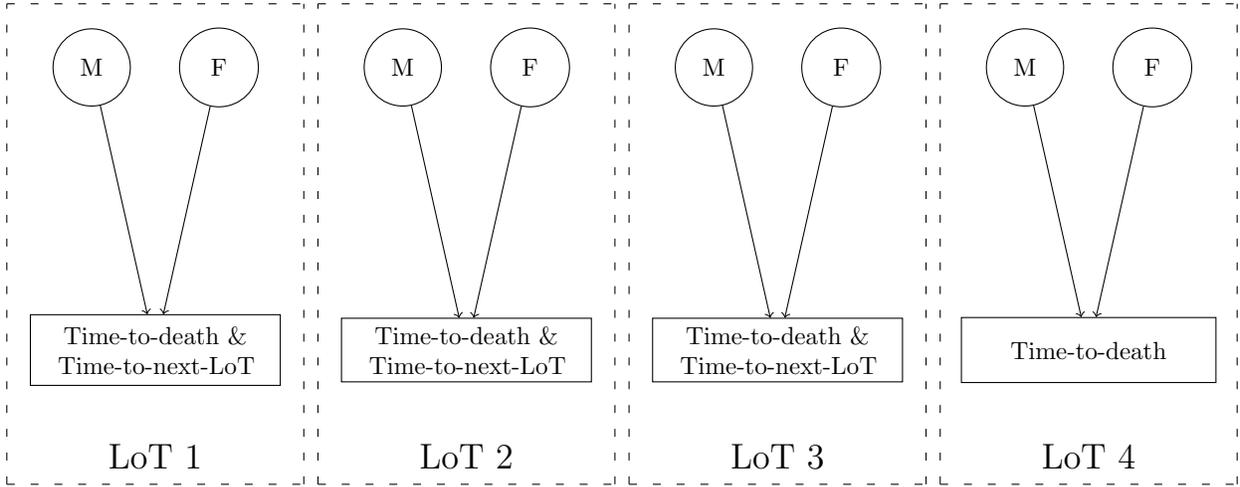
\begin{figure}[ht]
    \centering
        \begin{tikzpicture}[scale=0.94, transform shape]
        \tikzstyle{sty-01} = [rectangle, draw]
        \tikzstyle{sty-02} = [circle, draw]
        \node[sty-01, text width=3.3cm, minimum height=1cm] (n0) at (-9.2,-1.5) 
        {\quad Time-to-death \& \\ \;\, Time-to-next-LoT};
        \node[sty-02] (n1) at (-10.1,+2.5) 
        {\; M \;};
        \node[sty-02] (n3) at (-8.3,+2.5) 
        {\;\, F \;\,};
        \node[sty-01, text width=3.3cm] (n4) at (-4.8,-1.5) 
        {\quad Time-to-death \& \\ \;\, Time-to-next-LoT};
        \node[sty-02] (n5) at (-5.7,+2.5) 
        {\; M \;};
        \node[sty-02] (n7) at (-3.9,+2.5) 
        {\;\, F \;\,};
        \node[sty-01, text width=3.3cm] (n8) at (-0.4,-1.5) 
        {\quad Time-to-death \& \\ \;\, Time-to-next-LoT};
        \node[sty-02] (n9) at (-1.3,+2.5) 
        {\; M \;};
        \node[sty-02] (n11) at (0.5,+2.5) 
        {\;\, F \;\,};
        \node[sty-01, minimum width=3.6cm, minimum height=0.92cm] (n12) at (4,-1.5) 
        {Time-to-death};
        \node[sty-02] (n13) at (3.1,+2.5) 
        {\; M \;};
        \node[sty-02] (n15) at (4.9,+2.5) 
        {\;\, F \;\,};
        \draw[->] (n1) -> (n0);
        \draw[->] (n3) -> (n0);
        \draw[->] (n5) -> (n4);
        \draw[->] (n7) -> (n4);
        \draw[->] (n9) -> (n8);
        \draw[->] (n11) -> (n8);
        \draw[->] (n13) -> (n12);
        \draw[->] (n15) -> (n12);
        \draw (-9.2,-3.0) node {\Large LoT 1};
        \draw[loosely dashed] (-11.3,-3.4)--(-11.3,3.4);
        \draw[loosely dashed] (-11.3,-3.4)--(-7.1,-3.4);
        \draw[loosely dashed] (-11.3,3.4)--(-7.1,3.4);
        \draw[loosely dashed] (-7.1,-3.4)--(-7.1,3.4);
        \draw (-4.8,-3.0) node {\Large LoT 2};
        \draw[loosely dashed] (-6.9,-3.4)--(-6.9,3.4);
        \draw[loosely dashed] (-6.9,-3.4)--(-2.7,-3.4);
        \draw[loosely dashed] (-6.9,3.4)--(-2.7,3.4);
        \draw[loosely dashed] (-2.7,-3.4)--(-2.7,3.4);
        \draw (-0.4,-3.0) node {\Large LoT 3};
        \draw[loosely dashed] (-2.5,-3.4)--(-2.5,3.4);
        \draw[loosely dashed] (-2.5,-3.4)--(1.7,-3.4);
        \draw[loosely dashed] (-2.5,3.4)--(1.7,3.4);
        \draw[loosely dashed] (1.7,-3.4)--(1.7,3.4);
        \draw (4,-3.0) node {\Large LoT 4};
        \draw[loosely dashed] (1.9,-3.4)--(1.9,3.4);
        \draw[loosely dashed] (1.9,-3.4)--(6.1,-3.4);
        \draw[loosely dashed] (1.9,3.4)--(6.1,3.4);
        \draw[loosely dashed] (6.1,-3.4)--(6.1,3.4);
        \end{tikzpicture}
    \caption{Graphical representation of joint modelling for MM data based on a blockwise inferential scheme. Circle: longitudinal submodel. Rectangle: time-to-event submodel. M: M-spike, F: free light chains. LoT: line of therapy.} \label{fig:2StJM}
\end{figure}

A joint model of longitudinal and survival outcomes was developed to account for the complexity of the data and research questions \citep{rizopoulos2012}. Specifically, biomarkers are endogenous time-varying covariates, where their trajectories can be modelled through a suitable longitudinal submodel; time-to-death and time-to-next-LoT can be modelled with a competing risks submodel, where individual-level information is allowed to be shared across submodels. In addition, we considered a Bayesian approach due to the ease of incorporating it into hierarchical structures, quantifying uncertainty, and making dynamic predictions as new observations become available \citep{desmee2017, alsefri2020, kerioui2020}. Although a multistate approach is the natural first choice for modelling progressive transitions; in practice, we had convergence problems due to the complexity of the multistate likelihood function that includes simultaneously estimating the random effects from the biomarker submodels along the transitions/LoTs. Hence, we adopted a blockwise inferential scheme, which reduces the complexity of a multistate framework by considering independent blocks of competing risk specifications \citep{chen2024}. Figure~\ref{fig:2StJM} illustrates this strategy. Note that each LoT has its own joint model and each of these is independent of the others. This reduces the number of parameters to be estimated simultaneously and provides the possibility of running joint models in parallel. It is also worth highlighting that this strategy naturally considers a clock-reset specification \citep{kleinbaum2012} at the start of each LoT.

We describe in the following the step-by-step construction of our Bayesian joint model: its submodels and the setting for priors.

\subsection{Longitudinal Submodels} \label{subsec:biexponential}

We specify the longitudinal processes that model M-spike ($k=1$) and free light chains ($k=2$) biomarkers through a bi-exponential model \citep{stein2008,claret2009}. This class of model is quite popular in Pharmacometrics \citep{bruno2023,goncalves2024}, where, for example, tumour size dynamics exhibit nonlinear trajectories with potential initial decline (response to treatment) and flexibility to maintain decline, stabilise on a plateau, or grow over time (relapse, disease progression). These nonlinear behaviours are also expected for M-spike and FLC (see Web Figure~\ref{fig1}), which shows the posterior mean trajectories for both biomarkers in each LoT). Mathematically, such a model is given by
\begin{equation}
    y_{kli}(t) = B_{kli} \Big[ \exp\left\{ G_{kli} t \right\} + \exp\left\{ -D_{kli} t \right\} - 1 \Big] + \epsilon_{kli}(t), \label{eq:biexponential}
\end{equation}

\noindent where $y_{kli}(t)$ represents the observed value of biomarker $k=1,2$ in line of therapy (LoT) $l=1,2,3,4$ for patient $i=1,\ldots,n_{l}$ at time $t$ ($t=0$ indicates the therapy start time). $B_{kli}$, $G_{kli}$, and $D_{kli}$ are parameters that take only positive values and represent baseline (the biomarker value at $t=0$), growth rate, and decay rate, respectively, which are characteristics associated with the biomarker's longitudinal trajectory. The residual errors, $\epsilon_{kl1}(t),\ldots,\epsilon_{kln_{l}}(t)$, are assumed additive, independent and identically distributed as $\epsilon_{kli}(t) \sim \mbox{Normal}(0,\sigma_{kl}^{2})$.

We redefine the three parameters of the \eqref{eq:biexponential} as $B_{kli} = \exp\left\{\theta_{1kl} + b_{kli1}\right\}$, $G_{kli} = \exp\left\{\theta_{2kl} + b_{kli2}\right\}$, and $D_{kli} = \exp\left\{\theta_{3kl} + b_{kli3}\right\}$, where ${\bm \theta}_{kl}=(\theta_{1kl},\theta_{2kl},\theta_{3kl})^{\top}$ are population parameters while ${\bm b}_{kli}=(b_{kli1},b_{kli2},b_{kli3})^{\top}$ are random effects. In addition, we assume that ${\bm b}_{kli} \sim \mbox{Normal}({\bm 0},{\bm\Omega}_{kl})$, where ${\bm\Omega}_{kl}$ is an unstructured variance-covariance matrix.

Note that we opted to specify two univariate bi-exponential models (i.e., $k=1$ and $k=2$ in \eqref{eq:biexponential} are independent of each other) instead of a bivariate one. This simplification was made to overcome convergence issues associated with a common unstructured variance-covariance matrix for all random effects. Moreover, in medical practice, these biomarkers are expected to be complementary but not necessarily correlated \citep{tacchetti2017, gran2021}, which corroborates what we assumed.

\subsection{Competing Risks Submodels} \label{subsec:CR}

We model time-to-death ($v=1$) and time-to-next-LoT ($v=2$) through a competing risks model \citep{putter2007}, via a proportional cause-specific hazard specification \citep{putter2020}. We denote $T_{lvi}$ as the time from the start of LoT $l$ to the occurrence of event $v$ for patient $i$; $C_{li}$ indicates the censoring time for patient $i$ in LoT $l$; $\delta_{li}=0,1,2$ is an event indicator, where $\delta_{li}=0$ represents censoring for both events in LoT $l$, $\delta_{li}=1$ indicates that patient $i$ died in LoT $l$, and $\delta_{li}=2$ that patient $i$ transitioned to LoT $l+1$; and $T_{li} = \min\{T_{l1i},T_{l2i},C_{li}\}$ represents the observed event time for patient $i$ in LoT $l$. For a LoT $l$, we specify the hazard function of patient $i$ for event $v$ at time $t$ given by
\begin{equation}
     h_{lvi}(t) = h_{lv0}(t)\exp\left\{{\bm X}_{lvi}^{\top}{\bm \beta}_{lv} + \sum_{k=1}^{2}\big(\alpha_{klv1}B_{kli}^{\ast} + \alpha_{klv2}G_{kli}^{\ast} + \alpha_{klv3}D_{kli}^{\ast}\big) \right\}, \label{eq:CR}
\end{equation}

\noindent where $h_{lv0}(t)$ represents a baseline hazard function and is defined throughout this work as a Weibull hazard given by $h_{lv0}(t) = \phi_{lv} t^{\phi_{lv}-1}\exp\{\beta_{lv0}\}$, where $\phi_{lv}$ and $\beta_{lv0}$ are shape and log-scale parameters; ${\bm X}_{lvi}$ is a covariate vector with coefficients ${\bm \beta}_{lv}$; $B_{kli}^{\ast} = \log(B_{kli})$, $G_{kli}^{\ast} = \log(G_{kli})$, and $D_{kli}^{\ast} = \log(D_{kli})$ are the baseline, growth rate, and decay rate (in log scale) of biomarker $k$ in LoT $l$ for patient $i$, shared from the longitudinal submodel \eqref{eq:biexponential}, where $\alpha_{klv1}$, $\alpha_{klv2}$, and $\alpha_{klv3}$ have the role of measuring the strength of association between each characteristic of the biomarker trajectory and the risk for event $v$. As we adopted a cause-specific competing risks specification, the overall survival function of \eqref{eq:CR} for $v=1,2$ is defined as $S_{li}(t) = \exp\{-H_{l1i}(t) - H_{l2i}(t)\}$, where $H_{lvi}(t)$ is the cumulative hazard of $h_{lvi}(t)$. Note that for LoT $l=4$ we do not have a competing risks specification (see Figure~\ref{fig:2StJM}) but instead a proportional hazard model for time-to-death ($v=1$ only).

In preliminary analyses, we evaluated different parametric baseline hazard functions (exponential and Gompertz) and different shared terms (current value, only one of the parameters $B_{kli}^{\ast}$, $G_{kli}^{\ast}$, and $D_{kli}^{\ast}$, or pairs thereof), including dependence on previous LoT parameters, $B_{k(l-1)i}^{\ast}$, $G_{k(l-1)i}^{\ast}$, and $D_{k(l-1)i}^{\ast}$, which were not significant. The best specification was the one used in \eqref{eq:CR} (without previous LoT summaries) according to the leave-one-out cross-validation (LOO-CV) and the widely applicable information criterion (WAIC) \citep{vehtari2017}.

\subsection{Prior Elicitation} \label{subsec:priors}

We assume independent and weakly informative marginal prior distributions \citep{gelman2013}. For each biomarker $k=1,2$ and LoT $l=1,2,3,4$ using the longitudinal submodel \eqref{eq:biexponential}, the population parameters $\theta_{1kl}$, $\theta_{2kl}$, and $\theta_{3kl}$ follow Normal($0,10^2$) prior distributions, the residual error variance, $\sigma_{kl}^{2}$, follows a half-Cauchy($0,5$) prior distribution \citep{gelman2006}, and the random effects variance-covariance matrix ${\bm\Omega}_{kl}$ follows an inverse-Wishart($\mathbf{I}_{3},4$) prior distribution \citep{schuurman2016}, where $\mathbf{I}_{3}$ represents a $3 \times 3$ identity matrix. For each LoT $l=1,2,3,4$ and competing risk events $v=1,2$ using the survival submodel \eqref{eq:CR}, the regression coefficients ${\bm \beta}_{lv}$ (including the Weibull log-scale $\beta_{lv0}$) and the association parameters $\alpha_{klv1}$, $\alpha_{klv2}$, and $\alpha_{klv3}$ follow Normal($0,10^2$) prior distributions, and the Weibull shape parameter, $\phi_{lv}$, follows a half-Cauchy($0,1$) prior distribution \citep{rubio2018}. We previously investigated the sensitivity of our posterior distributions to more or less vague prior distributions, specifically Normal($0,100^{2}$) and half-Cauchy($0,15$). We concluded that our choice of prior distributions is weakly informative, since the results were equivalent across scenarios, differing only in computational time.

\section{Posterior Inference and Prediction} \label{sec:approaches}

\subsection{Joint Estimation (JE) Approach} \label{subsec:jmapp}

For each LoT $l=1,2,3,4$, we assume that longitudinal processes ${\bm y}_{1l}$ and ${\bm y}_{2l}$ are independent of each other and that they are conditionally independent of competing risk process ${\bm s}_{l}$ given the shared information $({\bm \theta}_{1l}, {\bm \theta}_{2l}, {\bm b}_{1l}, {\bm b}_{2l})$, where ${\bm b}_{kl} = ({\bm b}_{kl1},\ldots,{\bm b}_{kln_{l}})$ is the vector of all individual-level random effects of biomarker $k$ in LoT $l$. So, for a LoT $l$, the joint posterior distribution of all parameters is proportionally expressed as follows:
\begin{equation} 
\begin{aligned}
\pi({\bm \theta}_{1l},{\bm \theta}_{2l},{\bm \Psi}_{1l},{\bm \Psi}_{2l},{\bm \Phi}_{l},{\bm b}_{1l},{\bm b}_{2l} \mid {\mathcal D}_{l}) &\propto f_{1}({\bm y}_{1l} \mid {\bm \theta}_{1l}, {\bm b}_{1l}, {\bm \Psi}_{1l})f_{2}({\bm y}_{2l} \mid {\bm \theta}_{2l}, {\bm b}_{2l}, {\bm \Psi}_{2l}) \times \\ 
& \hspace{-5cm} \times f_{3}({\bm s}_{l} \mid {\bm \theta}_{1l}, {\bm \theta}_{2l}, {\bm b}_{1l}, {\bm b}_{2l}, {\bm \Phi}_{l}) g_{1}({\bm b}_{1l} \mid {\bm \Psi}_{1l}) g_{2}({\bm b}_{2l} \mid {\bm \Psi}_{2l}) \pi({\bm \theta}_{1l},{\bm \theta}_{2l},{\bm \Psi}_{1l},{\bm \Psi}_{2l},{\bm \Phi}_{l}), \label{eq:jmpostdist}
\end{aligned}
\end{equation}

\noindent where ${\mathcal D}_{l} = \{{\bm y}_{1li},{\bm y}_{2li},T_{li},\delta_{li},{\bm X}_{l \cdot i}; \, i=1,\ldots,n_{l}\}$ denotes data available (training set) for LoT $l$; ${\bm \Psi}_{1l}=({\bm \Omega}_{1l}, \sigma_{1l}^{2})$, ${\bm \Psi}_{2l}=({\bm \Omega}_{2l}, \sigma_{2l}^{2})$, and ${\bm \Phi}_{l}=({\bm \beta}_{l}, {\bm \alpha}_{1l}, {\bm \alpha}_{2l}, {\bm \phi}_{l})$; the density functions $f_{1}(\cdot)$, $f_{2}(\cdot)$, and $f_{3}(\cdot)$, derived from the longitudinal submodel \eqref{eq:biexponential} for biomarkers $k=1,2$ and the competing risks submodel \eqref{eq:CR}, are expressed as follows:
\begin{align} 
f_{k}({\bm y}_{kl} \mid \cdot) &= \prod_{i=1}^{n_{l}}\prod_{j=1}^{m_{kli}} \frac{1}{\sqrt{2\pi\sigma_{kl}^{2}}} \exp\left\{-\frac{1}{\sigma_{kl}^{2}}\big(y_{kli}(t_{klij})-\mu_{kli}(t_{klij})\big)^{2}\right\}, \text{ for } k=1,2, \label{eq:f1f2dens} \\
f_{3}({\bm s}_{l} \mid \cdot) &= \prod_{i=1}^{n_{l}} h_{l1i}(T_{li})^{1_{(\delta_{li}=1)}} \, h_{l2i}(T_{li})^{1_{(\delta_{li}=2)}}\exp\left\{-\int_{0}^{T_{li}}\big[h_{l1i}(t) + h_{l2i}(t)\big]\text{d}t \right\}, \label{eq:f3dens}
\end{align}

\noindent where $\mu_{kli}(t) = B_{kli} \big[ \exp\left\{ G_{kli} t \right\} + \exp\left\{ -D_{kli} t \right\} - 1 \big]$ is the true and unobserved trajectory of biomarker $k=1,2$ evaluated at time $t$, and $t_{klij} = 1,\ldots,m_{kli}$ represent the time points at which biomarker values ($k=1,2$) are recorded in LoT $l$ for patient $i$; $g_{1}(\cdot)$ and $g_{2}(\cdot)$ are Normal densities for random effects ${\bm b}_{1l}$ and ${\bm b}_{2l}$, respectively; and $\pi(\cdot)$ is the prior distribution specified as in Section~\ref{subsec:priors}.

\subsection{Corrected Two-Stage (TS) Approach} \label{subsec:tsapp}

Typically, calculating posterior distributions from joint models, such as in \eqref{eq:jmpostdist}, is computationally demanding and their complexity may also cause convergence problems \citep{mehdizadeh2021}. Both issues are avoided using a standard two-stage estimation, where the longitudinal submodel is first fitted, shared quantities are estimated and, in a second stage, inserted as known covariates into the survival submodel. However, several studies have shown that this strategy leads to biased results \citep{tsiatis2004, ye2008, sweeting2011, hickey2016, leiva2021}. Recently, \citet{alvares2023} proposed a corrected two-stage approach that mitigates such estimation biases. The authors developed this methodology in the context of exponential family distributions with linear predictors for a longitudinal outcome and a proportional hazard model. \citet{alvares2024a} also used a univariate time-to-event outcome but extrapolated the original proposal to a bi-exponential model (i.e., a nonlinear predictor) with a multiplicative error term. Here we extend this corrected two-stage approach to multiple bi-exponential models and competing risks survival outcomes. More specifically, in the first stage, we calculate the {\it maximum a posteriori} of the parameters of each longitudinal submodel in LoT $l$:
\begin{equation} 
\begin{aligned}
(\hat{\bm \theta}_{1l}, \hat{\bm \Psi}_{1l}) &= \argmax_{{\bm \theta}_{1l}, \, {\bm \Psi}_{1l}} f_{1}({\bm y}_{1l} \mid {\bm \theta}_{1l}, {\bm b}_{1l}, {\bm \Psi}_{1l}) g_{1}({\bm b}_{1l} \mid {\bm \Psi}_{1l})\pi({\bm \theta}_{1l},{\bm \Psi}_{1l}), \\
(\hat{\bm \theta}_{2l}, \hat{\bm \Psi}_{2l}) &= \argmax_{{\bm \theta}_{2l}, \, {\bm \Psi}_{2l}} f_{2}({\bm y}_{2l} \mid {\bm \theta}_{2l}, {\bm b}_{2l}, {\bm \Psi}_{2l}) g_{2}({\bm b}_{2l} \mid {\bm \Psi}_{2l})\pi({\bm \theta}_{2l},{\bm \Psi}_{2l}), \label{eq:1stage}
\end{aligned}
\end{equation}

\noindent where $f_{1}(\cdot)$, $f_{2}(\cdot)$, $g_{1}(\cdot)$, $g_{2}(\cdot)$, and $\pi(\cdot)$'s are specified as in Section~\ref{subsec:jmapp}. Note that estimation of random effects is not required, so one can theoretically use the marginalised likelihood function (i.e., the random effects can be integrated out). In the second stage, the posterior distribution of $({\bm \Phi}_{l},{\bm b}_{1l},{\bm b}_{2l})$ is approximated considering the full joint likelihood function given $(\hat{\bm \theta}_{1l},\hat{\bm \theta}_{2l},\hat{\bm \Psi}_{1l},\hat{\bm \Psi}_{2l})$:
\begin{equation} 
\begin{aligned}
\pi({\bm \Phi}_{l},{\bm b}_{1l},{\bm b}_{2l} \mid \hat{\bm \theta}_{1l},\hat{\bm \theta}_{2l},\hat{\bm \Psi}_{1l},\hat{\bm \Psi}_{2l},{\mathcal D}_{l}) &\propto f_{1}({\bm y}_{1l} \mid \hat{\bm \theta}_{1l}, {\bm b}_{1l}, \hat{\bm \Psi}_{1l}) f_{2}({\bm y}_{2l} \mid \hat{\bm \theta}_{2l}, {\bm b}_{2l}, \hat{\bm \Psi}_{2l}) \times \\ 
& \hspace{-5cm} \times f_{3}({\bm s}_{l} \mid \hat{\bm \theta}_{1l}, \hat{\bm \theta}_{2l}, {\bm b}_{1l}, {\bm b}_{2l}, {\bm \Phi}_{l}) g_{1}({\bm b}_{1l} \mid \hat{\bm \Psi}_{1l}) g_{2}({\bm b}_{2l} \mid \hat{\bm \Psi}_{2l}) \pi({\bm \Phi}_{l}), \label{eq:2stage}
\end{aligned}
\end{equation}

\noindent where $f_{1}(\cdot)$, $f_{2}(\cdot)$, $f_{3}(\cdot)$, $g_{1}(\cdot)$, $g_{2}(\cdot)$, and $\pi(\cdot)$ are specified as in Section~\ref{subsec:jmapp}. 

This corrected two-stage approach reduces the number of parameters to be estimated simultaneously and corrects estimation biases by allowing random effects to be calculated in the second stage considering the likelihood function of the joint model \eqref{eq:biexponential}-\eqref{eq:CR}. This strategy leads to a ``shared parameter compensation'', where the random effects calculated in the second stage help to correct the bias of the parameters $\hat{\bm \theta}_{1l}$ and $\hat{\bm \theta}_{2l}$, estimated in the first stage.

\subsection{Model Performance}

To assess the goodness-of-fit of the joint model \eqref{eq:biexponential}-\eqref{eq:CR} for MM data as well as to compare the equivalence of using JE and TS approaches, we used the test data set to evaluate the following items: individual weighted and Cox-Snell residuals \citep{desmee2017b}, time-dependent area under the receiver operating characteristic curve (AUC) using inverse probability of censoring weighting (IPCW) as a measure of discrimination \citep{blanche2015, blanche2019}, and calibration plot that assesses the agreement between predicted and observed risk \citep{paige2018, austin2022}. Individual weighted residuals are defined as $\text{IWRES}_{kli}(t) = \big(y_{kli}(t)-\hat{y}_{kli}(t)\big)/\hat{\sigma}_{kl}$, where $\hat{y}_{kli}(t)$ is the predicted value of biomarker $k$ in LoT $l$ for patient $i$ at time $t$ and $\hat{\sigma}_{kl}$ is the estimated standard deviation for the residual term (see Section~\ref{subsec:biexponential}). Cox-Snell residuals are defined as $\text{CS}_{li} = H_{l1i}(T_{i}) + H_{l2i}(T_{i})$, where $H_{lvi}(t)$ is the cumulative hazard of $h_{lvi}(t)$ for $v=1,2$. If the joint model properly fits the data, the Kaplan–Meier curve of $\text{CS}_{li}$ is expected to superimpose the survival curve of the unit exponential distribution \citep{rizopoulos2012}.

\subsection{Dynamic Risk Prediction}

One clinical interest is to obtain personalised risk predictions for death or change to next LoT based on the latest trajectory information about a given patient. As new observations of M-spike and/or free light chains biomarkers become available, the risk predictions should be dynamically updated \citep{taylor2005, yu2008, andrinopoulou2021}. More specifically, we would like to predict cumulative incidence probabilities for a patient $i^{\ast}$ in LoT $l$ who has provided us with a set of M-spike ($k=1$) and free light chains ($k=2$) longitudinal measurements, $y_{kli^{\ast}}(t) = \{y_{kli^{\ast}}(t_{k1}),\ldots,y_{kli^{\ast}}(t_{km_{k}}); \, 0 \leq t_{k1} < \ldots < t_{km_{k}} < t\}$ for $k=1,2$, and baseline characteristics, ${\bm X}_{l \cdot i^{\ast}}$. Given that no event occurred until $t$, we specify the conditional cumulative incidence function for patient $i^{\ast}$ in LoT $l$ at time $u>t$ as follows:
\begin{equation}
     F_{lvi^{\ast}}(u,t) = \mathbb{P}(T_{li^{\ast}} < u, \delta_{li^{\ast}} = v \mid T_{li^{\ast}} > t, {\mathcal D}_{li^{\ast}}, {\mathcal D}_{l}), \label{eq:CIF1}
\end{equation}

\noindent where $v=1$ and $v=2$ represent the competing events ``die'' and ``change to next LoT'', respectively, ${\mathcal D}_{li^{\ast}} = \{y_{1li^{\ast}}(t), y_{2li^{\ast}}(t), {\bm X}_{l \cdot i^{\ast}}\}$ denotes the data of patient $i^{\ast}$, and ${\mathcal D}_{l} = \{{\bm y}_{1li}, {\bm y}_{2li},$ $ T_{li}, \delta_{li}, {\bm X}_{l \cdot i}; \, i=1,\ldots,n_{l}\}$ denotes data available (training set) for LoT $l$.

Following \citet{rizopoulos2011}'s proposal adapted to a Bayesian and competing risks framework \citep{andrinopoulou2017}, Equation \eqref{eq:CIF1} can be rewritten as:
\begin{equation} 
\begin{aligned}
     F_{lvi^{\ast}}(u,t) &= \int \mathbb{P}(T_{li^{\ast}} < u, \delta_{li^{\ast}} = v \mid T_{li^{\ast}} > t, {\bm \Theta}_{l}, {\bm b}_{li^{\ast}}) \times \\
     & \hspace{3cm} \times \pi({\bm \Theta}_{l} \mid {\mathcal D}_{l}) \pi({\bm b}_{li^{\ast}} \mid T_{li^{\ast}} > t, {\mathcal D}_{li^{\ast}}, {\bm \Theta}_{l}) \, \text{d}({\bm \Theta}_{l},{\bm b}_{li^{\ast}}), \label{eq:CIF2}
\end{aligned}
\end{equation}

\noindent where ${\bm \Theta}_{l} = ({\bm \theta}_{1l},{\bm \theta}_{2l},{\bm \Psi}_{1l},{\bm \Psi}_{2l},{\bm \Phi}_{l})$ and ${\bm b}_{li^{\ast}} = ({\bm b}_{1li^{\ast}},{\bm b}_{2li^{\ast}})$. The second term of the integrand \eqref{eq:CIF2}, $\pi({\bm \Theta}_{l} \mid {\mathcal D}_{l})$, is the posterior distribution of ${\bm \Theta}_{l}$ from the joint model \eqref{eq:biexponential}-\eqref{eq:CR} using the training set ${\mathcal D}_{l}$ in LoT $l$. The third term of the integrand \eqref{eq:CIF2}, $\pi({\bm b}_{li^{\ast}} \mid T_{li^{\ast}} > t, {\mathcal D}_{li^{\ast}}, {\bm \Theta}_{l})$, is the conditional posterior distribution of the random effects for patient $i^{\ast}$ in LoT $l$ given their observation history and the parameter vector ${\bm \Theta}_{l}$. Finally, the first term of the integrand \eqref{eq:CIF2} can be rewritten as:
\begin{equation} 
\begin{aligned}
     \mathbb{P}(T_{li^{\ast}} < u, \delta_{li^{\ast}} = v \mid T_{li^{\ast}} > t, {\bm \Theta}_{l}, {\bm b}_{li^{\ast}}) & = \frac{\mathbb{P}(t < T_{li^{\ast}} < u, \delta_{li^{\ast}} = v \mid {\bm \Theta}_{l}, {\bm b}_{li^{\ast}})}{\mathbb{P}(T_{li^{\ast}} > t \mid {\bm \Theta}_{l}, {\bm b}_{li^{\ast}})} \\
     & = \frac{\text{CIF}_{lvi^{\ast}}(u, t)}{S_{li^{\ast}}(t)}, \label{eq:CIF3}
\end{aligned}
\end{equation}

\noindent where $S_{li^{\ast}}(t)$ denotes the overall survival function (see Section~\ref{subsec:CR}) and $\text{CIF}_{lvi^{\ast}}(u, t) = \int_{t}^{u} h_{lvi^{\ast}}(s)S_{li^{\ast}}(s)\text{d}s$ is the cumulative incidence function for event $v$ from $t$ to $u$. Hence, an estimate of $F_{lvi^{\ast}}(u,t)$ can be obtained using the following Monte Carlo simulation scheme:

\begin{enumerate}
    \item[(I)] Draw ${\bm \Theta}_{l}^{(j)}$ from the MCMC sample of the posterior distribution $\pi({\bm \Theta}_{l} \mid {\mathcal D}_{l})$.
    \item[(II)] Draw ${\bm b}_{li^{\ast}}^{(j)}$ from $\pi({\bm b}_{li^{\ast}} \mid T_{li^{\ast}} > t, {\mathcal D}_{li^{\ast}}, {\bm \Theta}_{l}^{(j)})$.
    \item[(III)] Compute $F_{lvi^{\ast}}^{(j)}(u,t) = \displaystyle \frac{\text{CIF}_{lvi^{\ast}}(u, t \mid {\bm \Theta}_{l}^{(j)}, {\bm b}_{li^{\ast}}^{(j)})}{S_{li^{\ast}}(t \mid {\bm \Theta}_{l}^{(j)}, {\bm b}_{li^{\ast}}^{(j)})}$.
\end{enumerate}

Steps (I)-(III) are repeated for $j=1,\ldots,J$, where $J$ denotes the number of Monte Carlo samples. We then can calculate a point estimate of $F_{lvi^{\ast}}(u,t)$ by averaging over $F_{lvi^{\ast}}^{(1)}(u,t),\ldots,F_{lvi^{\ast}}^{(J)}(u,t)$. Moreover, a 95\% credible interval can be obtained using the Monte Carlo sample percentiles. Note that MCMC samples from steps (I) and (II) also allow us to update the estimate of the biomarker's trajectory using the longitudinal submodel \eqref{eq:biexponential} \citep{papageorgiou2019}.

Using our corrected two-stage approach requires minor modifications to step (I). First, the parameters of the longitudinal submodels, $({\bm \theta}_{1l},{\bm \theta}_{2l},{\bm \Psi}_{1l},\bm \Psi_{2l})$, are not resampled with the inclusion of biomarker measurements from patient $i^{\ast}$, i.e., such parameters are fixed at their respective estimated values (see Equation~\eqref{eq:1stage}), but step (I) is still applied to the survival model parameters, ${\bm \Phi}_{l}$. So, we redefine ${\bm \Theta}_{l}^{(j)}$ as $(\hat{\bm \theta}_{1l},\hat{\bm \theta}_{2l},\hat{\bm \Psi}_{1l},\hat{\bm \Psi}_{2l},{\bm \Phi}_{l}^{(j)})$, where ${\bm \Phi}_{l}^{(j)}$ is drawn from the MCMC sample of the posterior distribution $\pi({\bm \Phi}_{l} \mid {\mathcal D}_{l})$.

\section{Results} \label{sec:results}

All models were implemented in Stan using the \texttt{rstan} package version 2.26.23 \citep{rstan2023} from the R language version 4.3.1 \citep{rlang2023}. All codes are available at \url{www.github.com/daniloalvares/BJM-MBiExp-CR}. Warm-up and MCMC samples were specified as the minimum number of iterations, collected from three independent chains, for convergence (Gelman-Rubin statistic, R-hat $<$ 1.05) and efficiency (effective sample size, neff $>$ 100) to be achieved \citep{vehtari2021}. Then, JE and the first stage of TS used 4000 posterior samples after 1000 warm-up iterations, while the second stage of TS was run with a warm-up of 500 and then 1000 posterior samples. Once convergence and efficiency were reached, the three chains were pooled to estimate the posterior distributions for each parameter. Parameters were assessed using the posterior mean and the 95\% credible interval based on the 2.5th and 97.5th percentiles.

\subsection{Comparison of JE vs TS Model Performance}

We evaluated the performance of the joint model \eqref{eq:biexponential}-\eqref{eq:CR} considering the test data set using joint estimation (JE) and corrected two-stage (TS) approaches. In summary, JE and TS achieved similar and satisfactory performance in all metrics, especially before LoT 4. For longitudinal submodel fit, IWRES suggested that a bi-exponential specification was suitable for modelling M-spike and FLC along LoTs using JE or TS (see Web Figure~\ref{fig2}). For survival submodel fit, Cox-Snell residuals indicated a better fit using TS in LoTs 1, 2 \& 3 (see Web Figure~\ref{fig3}). For LoT 4, both estimation methods showed poor goodness-of-fit, with a small advantage in favor of JE, which has its curve closest to a unit exponential survival model (theoretical distribution). We hypothesise that many patients die or are censored after LoT 4 (362 cases, i.e., 50.5\% of patients in LoT 4) and so biomarker trajectories (in LoT 4) are not informative for predicting time-to-death for such patients. For prediction evaluation, time-dependent AUCs showed similar discrimination results between two estimation methods, with the exception of LoT 3 for the time-to-death submodel (see Figure~\ref{fig4}). We also find that at later landmarks, it is harder to discriminate between them. Finally, calibration plots assessed the agreement between predicted and observed risk using JE and TS, where highly comparable results were observed (see Figure~\ref{fig5a} and Web Figure~\ref{fig5b} for a comparison between JS and TS by decile and tercile of predicted 1-year risk, respectively). Note that division by decile makes the predicted risk sensitive to small sample size. For example, in Figure~\ref{fig5a}, the calibration plot for time-to-death in LoTs 3 and 4 looks like overpredicting the risk. However, when considering fewer risk subgroups (e.g. tercile), the predictions are more robust (as the sample size is larger), as can be seen in Web Figure~\ref{fig5b}.

\subsection{Comparison of JE vs TS Model Fit}

While longitudinal parameter estimates obtained with either JE or TS approach were in the same order of magnitude (see Web Tables~\ref{table:postbiexpJE} and \ref{table:postbiexpTS}, some notable differences were observed in the M-spike and FLC time dynamics as illustrated in Web Figure~\ref{fig1}. We highlight that the (population) longitudinal parameters estimated in the first stage are potentially biased in missing not at random (MNAR) censoring scenarios. Nevertheless, the role of the population parameters is to provide sufficient information on population characteristics (not necessarily unbiased) while the random effects re-estimated in the second stage effectively correct the bias of the individual parameter estimates.

In all cases, JE presented wider credible intervals, which is an expected result, since simultaneous inference considers more sources of uncertainty. Furthermore, the posterior mean trajectories using JE were above those estimated with TS, which is a finding of the influence of time-to-event information when estimating the longitudinal parameters. For the survival submodels, JE and TS produced similar results for both time-to-death (see Web Tables~\ref{table:postdeathJE} and \ref{table:postdeathTS}) and time-to-next-LoT (see Web Tables~\ref{table:postnextJE} and \ref{table:postnextTS}).

We highlight below common findings across the four LoTs for each clinical outcome. For time-to-death, patients who are older, with ECOG status 2$^{+}$, low platelet count, high initial M-spike value, or high initial FLC value (in each LoT) have a higher risk of death. For time-to-next-LoT, patients who are non-Hispanic white, young, with low immunoglobulin G levels, short time spent in the previous LoT, high initial M-spike value, high initial FLC value, high growth rate for M-spike, or high growth rate for FLC (in each LoT) have a higher risk of starting a next LoT.

\subsection{Dynamic Risk Prediction Examples}

To illustrate individual dynamic predictions, i.e., predictions that are updated once more information becomes available for the patient, we randomly selected two patients, named A and B, whose baseline characteristics are presented in Web Table~\ref{table:dynpredpatients}.

One-year dynamic predictions by using different lengths of histories of the respective patient ($t_1=$ first longitudinal observation, $t_{2}=$ up to 50\% of longitudinal observations, and $t_{3}=$ up to the last longitudinal observation) are shown using TS (Figures~\ref{fig6TS} and \ref{fig7TS}) and JE (Web Figures~\ref{fig6JE} and \ref{fig7JE}) approaches. Comparatively, TS or JE are largely similar.

For patient A (Figure~\ref{fig6TS}), M-spike and FLC decline over the first two LoTs, which could explain the fact that competing events have similar probabilities of occurrence and less than 50\%. This behaviour is maintained for the first two landmark time points in LoT 3, until both biomarkers begin to increase their values (M-spike more quickly) and then the probability of the patient requiring a change of LoT also increases. In LoT 4, biomarkers resume their downward trend, which helps reduce the probability of death for patient A over time.

For patient B (Figure~\ref{fig7TS}), M-spike value drops abruptly between the first and second measurement in LoT 1, while FLC values decrease slightly. Such behaviours reduce the probability of starting a next LoT and keep both competing events with a probability of occurrence between 20\% and 30\%. Note that the initiation of LoT 2 is started approximately after one year without longitudinal follow-up of the patient. In LoT 2, although both biomarkers present significant reductions, the risk of changing to LoT 3 remains around 50\%. Patient B has low biomarker values over the first few months on line 3 of therapy, but her last M-spike and FLC measurements rise rapidly, which likely explains the increase at risk of starting a next LoT. In LoT 4, M-spike values remain more or less stable around 10 g/L while FLC values increase over time. Both biomarker trajectories combined with the baseline characteristics of patient B increase her risk of death.

\begin{landscape} \thispagestyle{empty}

\begin{figure}[htp!] \vspace{-1.2cm}
  \centering
  \includegraphics[width=1.4\textwidth]{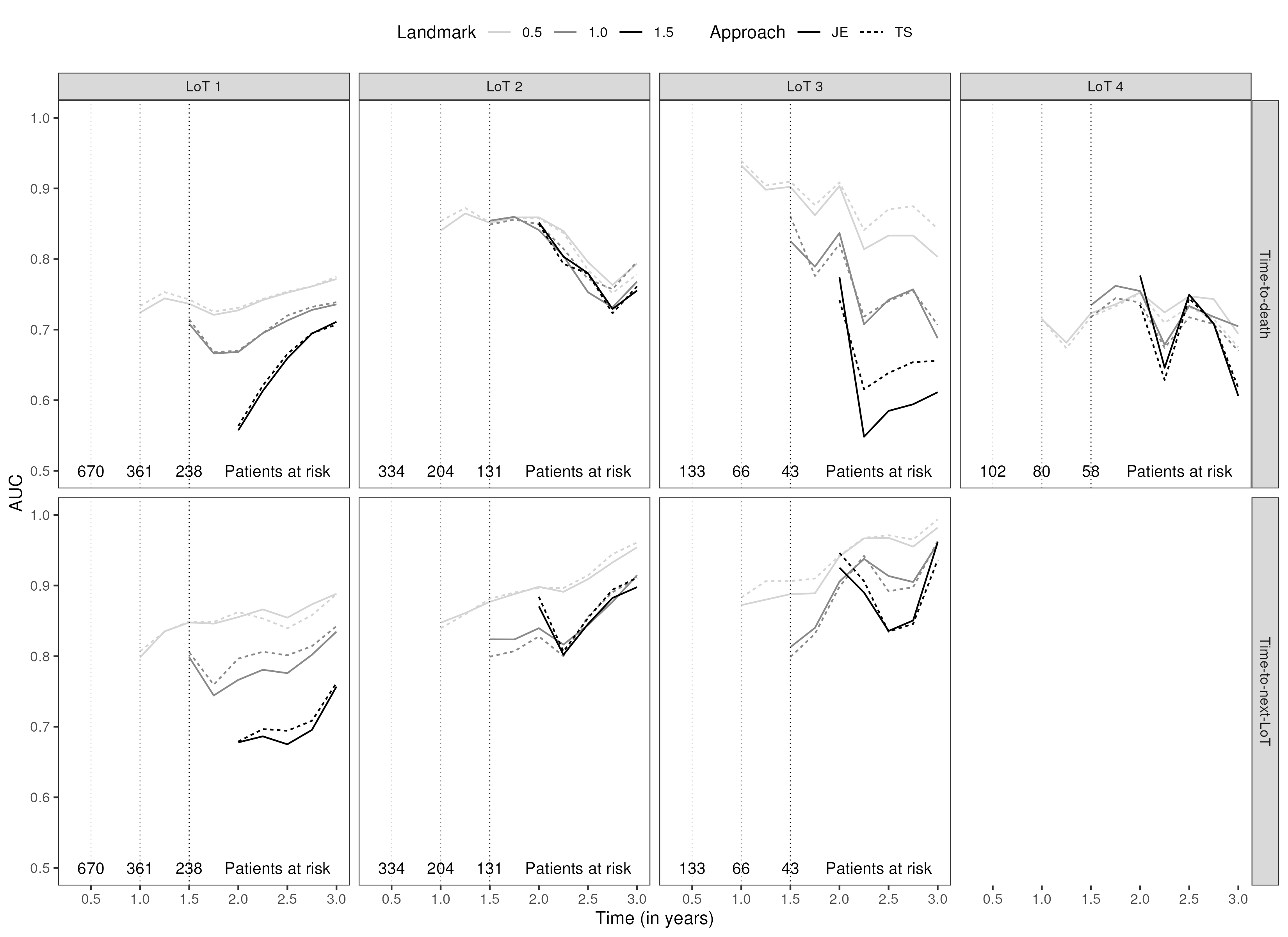}
  \caption{Time-dependent AUC from 6 months after landmark times at $0.5$, $1.0$, and $1.5$ years (from lightest to darkest respectively) for each time-to-event considering the test data set by line of therapy (LoT) using joint estimation (JE, solid line) and corrected two-stage (TS, dashed line) approaches.}
  \label{fig4}
\end{figure}

\end{landscape}

\begin{landscape} \thispagestyle{empty}

\begin{figure}[htp!] \vspace{-1.2cm}
  \centering
  \includegraphics[width=1.4\textwidth]{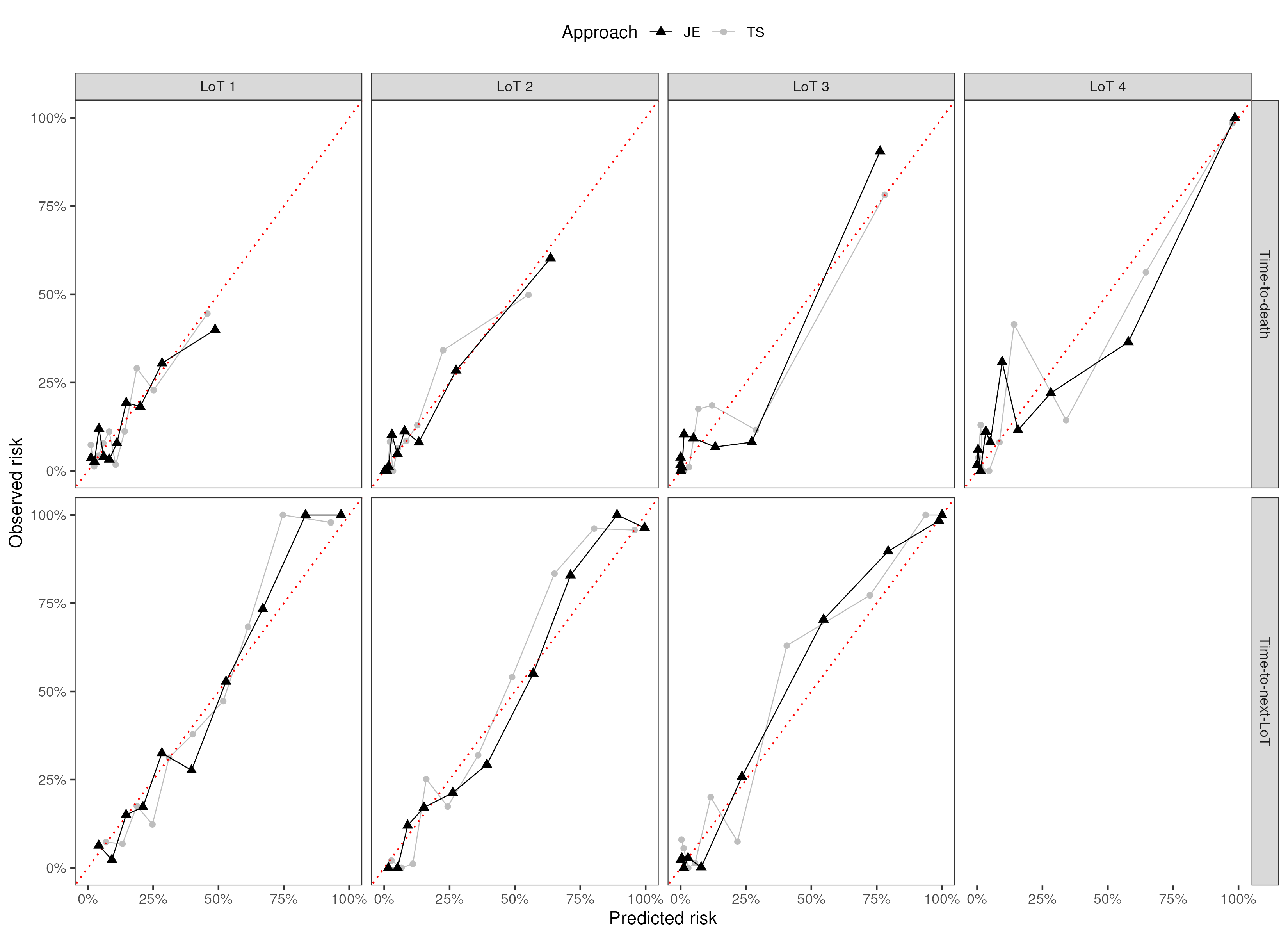}
  \caption{Calibration plots by decile of predicted 1-year risk from the survival submodel with 6 months of landmark time for each time-to-event considering the test data set by line of therapy (LoT) using joint estimation (JE, black triangle) and corrected two-stage (TS, gray circle) approaches.}
  \label{fig5a}
\end{figure}

\end{landscape}

\begin{figure}[H] \vspace{-1cm}
  \centering
  \includegraphics[width=0.9\textwidth]{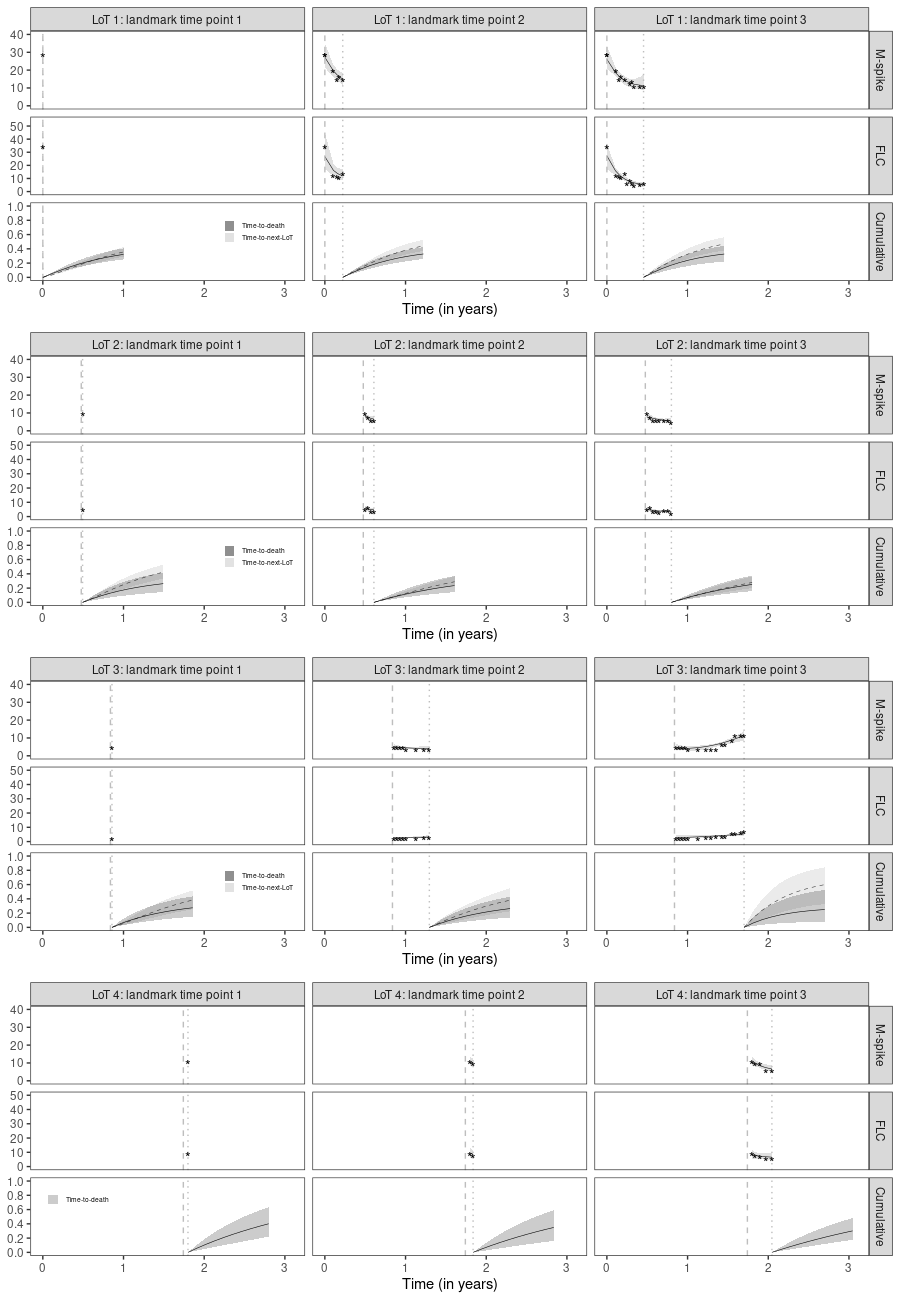}
  \caption{One-year dynamic predictions from three landmark time points (vertical dotted lines) in each line of therapy (LoT) using the corrected two-stage (TS) approach for patient A. Vertical dashed lines indicate LoT initiation times. For ``M-spike'' and ``FLC'' rows, stars represent biomarker observed values with their respective posterior mean trajectory (solid lines) and 95\% credible intervals (gray shadow). For ``Cumulative'' rows, solid and dashed lines are posterior means (with their respective 95\% credible intervals) of cumulative incidence functions (LoTs 1, 2 \& 3) or distribution functions (LoT 4) for time-to-death and time-to-next-LoT, respectively.}
  \label{fig6TS}
\end{figure}

\begin{figure}[H] \vspace{-1cm}
  \centering
  \includegraphics[width=0.9\textwidth]{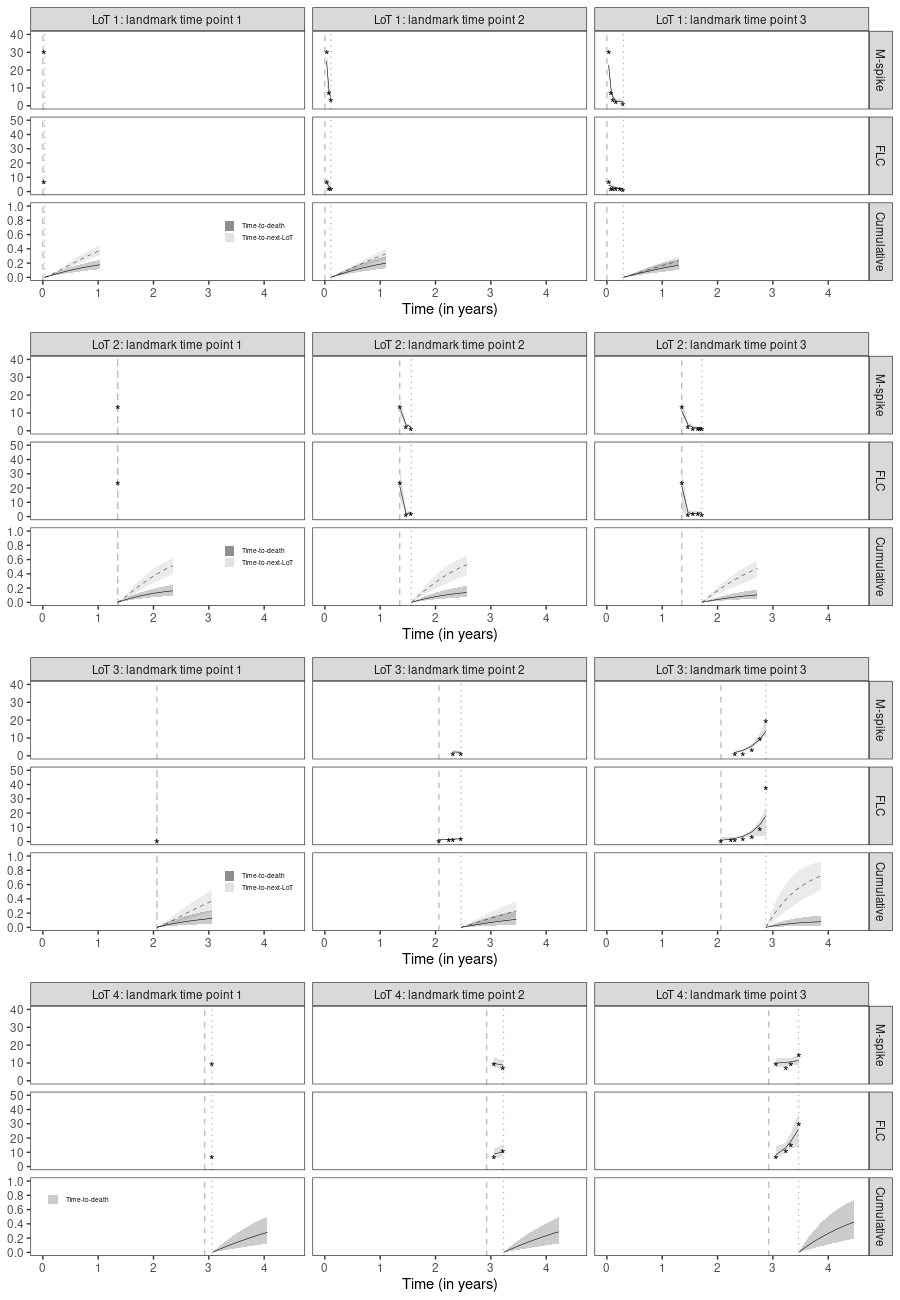}
  \caption{One-year dynamic predictions from three landmark time points (vertical dotted lines) in each line of therapy (LoT) using the corrected two-stage (TS) approach for patient B. Vertical dashed lines indicate LoT initiation times. For ``M-spike'' and ``FLC'' rows, stars represent biomarker observed values with their respective posterior mean trajectory (solid lines) and 95\% credible intervals (gray shadow). For ``Cumulative'' rows, solid and dashed lines are posterior means (with their respective 95\% credible intervals) of cumulative incidence functions (LoTs 1, 2 \& 3) or distribution functions (LoT 4) for time-to-death and time-to-next-LoT, respectively.}
  \label{fig7TS}
\end{figure}

\section{Discussion} \label{sec:discuss}

We have proposed a new Bayesian joint model for MM data that captures the dynamics of multiple biomarkers (M-spike and free light chains) through nonlinear mixed-effect submodels and shares characteristics of such biomarkers with a competing risks submodel, where the events of interest are death and change to next line of therapy. As an alternative to joint estimation (JE), we have extended the corrected two-stage (TS) approach proposed by \citet{alvares2023} to this more complex joint model.

Model performance was evaluated using analyses of residuals and predictive performance metrics. Longitudinal and Cox-Snell residuals demonstrated satisfactory model fit for both JE and TS approaches. In addition, time-dependent AUCs and calibration plots have shown equally good predictive performance for both approaches. Moreover, posterior inferences presented similar conclusions regardless of the estimation approach used, but TS has required much less computational effort (46.5\% reduction in processing time with 15.4 hours using TS vs 28.8 hours using JE).

We have revisited the dynamic prediction scheme for joint models introduced by \citet{rizopoulos2011} and discussed the minor modifications required to use it with the corrected two-stage estimation. We have also illustrated the applicability of dynamic risk predictions as an essential tool to better understand prognostic factors and their short-term and long-term impacts on patient journeys. For example, such predictions may inform optimal treatment strategies by risk status, support clinical decision-making at the point of care, and provide insights for clinical trial designs.

In conclusion, we have contributed to both hematology and statistical modelling literature with a new Bayesian predictive joint model that incorporates dynamic information from biomarker trajectories into a competing risks submodel. In terms of estimation methods, we prioritise JE but advocate TS as a robust approximation to JE when there are convergence issues or high computational time. \citet{alvares2024a} discussed the possibility of combining them, where TS can be used for variable selection and then JE estimates the final model. Our joint model can be extended in different directions, such as multiple change-points in more complex longitudinal data and multistate specifications. Furthermore, we would like to discuss some issues not covered in this work but which may be motivation for future research. For simplicity, we have handled missing data using the mean-value imputation, but other more sophisticated approaches could be explored, such as multiple imputation and machine learning techniques \citep{emmanuel2021}.
We have opted to use the well-known and popular dynamic prediction strategy proposed by \citet{rizopoulos2011}, but existing literature also provides alternatives to dynamically update predictions when new longitudinal measurements become available, such as sequential Monte Carlo methods \citep{alvares2021}. From a computational optimisation perspective, integrated nested Laplace approximation \citep{rue2009} is an alternative to speed up the inferential process for both JE and TS approaches. However, to the best of our knowledge, there are no implementations available yet for a joint model of multiple bi-exponential longitudinal submodels sharing their random effects with a competing risks submodel \citep{niekerk2021, niekerk2023, alvares2024b}. Another option, especially for big datasets, is to use the approach proposed by \citet{afonso2023}, where data is divided into subsamples, joint models are fitted to each of them in parallel, and then a consensus distribution strategy is applied to unify the results. From a clinical perspective, the treatment regimen in each LoT can also be considered as a predictor. However, in our study, when incorporating such a regimen into the survival submodel there were no improvements in predictions, so we dropped it in the main analysis. It is also worth mentioning that starting a new LoT (one of the competing events) is a human decision that presumably involves multiple factors not considered in this work, such as comorbidities, aggressive clinical features, prior toxicities, treatment guidelines, etc. \citep{laubach2016, mikhael2019, dimopoulos2021}. Hence, we hope that in the future such information will be available to be incorporated into the modelling process.

\section*{Data availability statement}

For eligible studies qualified researchers may request access to individual patient-level clinical data through a data request platform. For up-to-date details on Roche's Global Policy on the Sharing of Clinical Information and how to request access to related clinical study documents, see the website (\url{https://go.roche.com/data_sharing}). Anonymised records for individual patients across more than one data source external to Roche cannot, and should not, be linked due to a potential increase in risk of patient re-identification. The data that support the findings of this study were originated by and are the property of Flatiron Health, Inc., which has restrictions prohibiting the authors from making the data set publicly available. Requests for data sharing by license or by permission for the specific purpose of replicating results in this manuscript can be submitted to PublicationsDataAccess@flatiron.com. The data are subject to a license agreement with Flatiron Health to protect patient privacy and ensure compliance with measures necessary to reduce the risk of re-identification. For example, the data necessary to replicate the study include numerous specific dates, including visit dates (i.e., laboratory or examination dates), treatment start and stop dates, and month of death, as well as laboratory test results. Other measures to maintain de-identification without contractual agreements in place are not feasible due to the study question, methods used, and data elements required.

\section*{Acknowledgements}

D.A. and J.K.B. were supported by the U.K. Medical Research Council grant MC\_UU\_00002/5 and the collaboration grant jointly funded by Roche and the University of Cambridge. We thank the following colleagues for helpful discussions during the research and review stage of this work: Vallari Shah, Mellissa Williamson, Sarwar Mozumder, Madlaina Breuleux, Pascal Chanu, and Chris Harbron. For the purpose of open access, the author has applied a Creative Commons Attribution (CC BY) license to any Author Accepted Manuscript version arising from this submission.

\bibliographystyle{biom}
\bibliography{refs}


\captionsetup[figure]{name=Web Figure}
\captionsetup[table]{name=Web Table}
\setcounter{figure}{0}
\setcounter{table}{0}

\begin{center}
    {\LARGE \textbf{SUPPLEMENTARY MATERIAL}}
\end{center}

\appendix

\section*{Appendix A. Web Tables}

\begin{table}[htp!]
\caption{Summary (number of cases and percentage) of categorical variables by line of therapy (LoT). In terms of modelling, the first category of each variable is defined as the reference category.}
\label{table:catdescript}
\centering
\begin{tabular}{lcccc}
\hline
{\bf Baseline variable}  & {\bf LoT 1}   & {\bf LoT 2}   & {\bf LoT 3}  & {\bf LoT 4}  \\
\hline
Sex                &               &               &              &              \\
~~Male               & 2981 (54.3\%) & 1536 (55.4\%) & 756 (54.0\%) & 395 (55.2\%) \\
~~Female             & 2509 (45.7\%) & 1239 (44.6\%) & 645 (46.0\%) & 321 (44.8\%) \\
\hline
Ethnicity          &               &               &              &              \\
~~Non-Hispanic white & 3063 (55.8\%) & 1613 (58.1\%) & 838 (59.8\%) & 437 (61.0\%) \\
~~Non-Hispanic black & 906 (16.5\%)  & 445 (16.0\%)  & 217 (15.5\%) & 109 (15.2\%) \\
~~Other              & 960 (17.5\%)  & 486 (17.5\%)  & 239 (17.1\%) & 114 (15.9\%) \\
~~Not reported       & 561 (10.2\%)  & 231 (8.3\%)   & 107 (7.6\%)  & 56 (7.8\%)   \\
\hline
ECOG               &               &               &              &              \\
~~0                  & 1166 (21.2\%) & 597 (21.5\%)  & 303 (21.6\%) & 142 (19.8\%) \\
~~1                  & 1382 (25.2\%) & 694 (25.0\%)  & 358 (25.6\%) & 190 (26.5\%) \\
~~2$^{+}$            & 781 (14.2\%)  & 361 (13.0\%)  & 163 (11.6\%) & 77 (10.8\%)  \\
~~Not reported       & 2161 (39.4\%) & 1123 (40.5\%) & 577 (41.2\%) & 307 (42.9\%) \\
\hline
ISS                &               &               &              &              \\
~~Stage I            & 1144 (20.8\%) & 581 (20.9\%)  & 282 (20.1\%) & 141 (19.7\%) \\
~~Stage II           & 1129 (20.6\%) & 588 (21.2\%)  & 299 (21.3\%) & 142 (19.8\%) \\
~~Stage III          & 1141 (20.8\%) & 630 (22.7\%)  & 337 (24.1\%) & 192 (26.8\%) \\
~~Not reported       & 2076 (37.8\%) & 976 (35.2\%)  & 483 (34.5\%) & 241 (33.7\%) \\
\hline
\end{tabular}
\end{table}

\pagebreak

\begin{table}[htp!]
\caption{Summary of continuous variables in their original scales (Initial) and after log transformation$^{1}$\tnote{1}, standardisation, and imputation$^{2}$\tnote{2} (Final). NA represents the number of missing observations before imputation.}
\label{table:contdescript}
\centering
\begin{threeparttable}
\begin{tabular}{cccccccc}
\hline
{\bf Baseline variable}      & {\bf Data} & {\bf Mean} & {\bf SD}\tnote{3} & {\bf Median} & {\bf Min} & {\bf Max}  &  {\bf NA (\%)} \\
\hline
Age                          & Initial &  67.92 &  10.32 &  69.00 &    24.00 &   84.00 &   --   \\
(years)                      & Final   &   0.00 &   1.00 &   0.17 &  -6.18 &    1.36 &   --   \\
\hline
Albumin                      & Initial &  36.12 &  12.02 &  38.00 &     0.04 &  519.00 &  1512 (28) \\
(serum, g/L)                 & Final   &   0.00 &   0.85 &   0.03 & -10.95 &    5.46 &   --   \\
\hline
B2M                          & Initial &  10.01 & 103.88 &   4.00 &     0.20 & 3800.00 &  3041 (55) \\
(serum, mg/L)                & Final   &   0.00 &   0.67 &   0.00 &  -3.75 &    9.22 &   --   \\
                             \hline
Creatinine                   & Initial &   1.51 &   1.85 &   1.10 &     0.40 &   83.00 &  1546 (28) \\
(serum, mg/dL)               & Final   &   0.00 &   0.85 &   0.00 &  -1.95 &    8.09 &   --   \\
                             \hline
Hemoglobin                   & Initial &  10.78 &   2.17 &  10.70 &     3.10 &   22.50 &   736 (13) \\
(g/dL)                       & Final   &   0.00 &   0.93 &   0.00 &  -5.87 &    3.67 &   --   \\
                             \hline
LDH                          & Initial & 230.25 & 195.33 & 179.00 &    50.00 & 3182.00 &  3426 (62) \\
(serum, U/L)                 & Final   &   0.00 &   0.61 &   0.00 &  -2.72 &    5.53 &   --   \\
                             \hline
Lymphocyte                   & Initial &   1.82 &   1.23 &   1.60 &     0.00 &   32.70 &  1614 (29) \\
(count, $\times 10^{9}$/L)   & Final   &   0.00 &   0.84 &   0.00 &  -5.87 &    6.12 &   --   \\
                             \hline
Neutrophil                   & Initial &   5.93 &  87.97 &   3.50 &     0.00 & 4758.00 &  2252 (41) \\
(count, $\times 10^{9}$/L)   & Final   &   0.00 &   0.77 &   0.00 &  -6.32 &   12.75 &   --   \\
                             \hline
Platelet                     & Initial & 227.11 &  93.23 & 217.00 &     0.00 &  921.00 &  1635 (30) \\
(count, $\times 10^{9}$/L)   & Final   &   0.00 &   0.84 &   0.00 & -17.22 &    3.35 &   --   \\
                             \hline
IgA                          & Initial &   7.04 &  14.81 &   0.67 &     0.00 &  136.00 &  2663 (49) \\
(serum, g/L)                 & Final   &   0.00 &   0.72 &   0.00 &  -1.52 &    2.71 &   --   \\
                             \hline
IgG                          & Initial &  26.39 &  25.36 &  17.09 &     0.40 &  139.01 &  2544 (46) \\
(serum, g/L)                 & Final   &   0.00 &   0.73 &   0.00 &  -3.01 &    1.94 &   --   \\
                             \hline
IgM                          & Initial &   0.51 &   3.11 &   0.24 &     0.00 &   97.43 &  3105 (57) \\
(serum, g/L)                 & Final   &   0.00 &   0.66 &   0.00 &  -1.95 &    8.49 &   --   \\
\hline
\end{tabular}
  \begin{tablenotes}
    \item[1]$\log(x+0.1)$ to avoid numerical problems when $x=0$.
    \item[2]Mean-value imputation strategy for missing values \citep{donders2006}, i.e., we replace them with zeros (after standardisation, the mean of each variable is zero).
    \item[3]After standardisation, the standard deviation is equal to 1, but the imputation concentrates more values at zero and consequently reduces such standard deviation.
  \end{tablenotes}
\end{threeparttable}
\end{table}

\pagebreak

\begin{landscape} \thispagestyle{empty}

\begin{table}[htp!]
\caption{Posterior summary (mean and 95\% credible interval) of the bi-exponential submodel parameters for each biomarker by line of therapy (LoT) using the joint estimation (JE) approach. Statistically significant variables are shown in bold, except for variance parameters.} \label{table:postbiexpJE}
\centering
\begin{tabular}{cccccc}
\hline
{\bf Interpretation} & {\bf Parameter}  & {\bf LoT $\bm{l=1}$} &  {\bf LoT $\bm{l=2}$} &  {\bf LoT $\bm{l=3}$} &  {\bf LoT $\bm{l=4}$} \\ 
\hline
\multicolumn{6}{c}{M-spike ($k=1$)} \\
\hline  
Baseline        & $\exp\{\theta_{1kl}\}$ & {\bf 17.086} (16.405, 17.797) &  {\bf 7.765}  (7.347, 8.232) &  {\bf 7.382}  (6.780, 8.057) &  {\bf 7.950}   (7.030, 8.971)  \\ 
Growth          & $\exp\{\theta_{2kl}\}$ &  {\bf 0.246}   (0.234, 0.259) &  {\bf 0.293}  (0.257, 0.332) &  {\bf 0.377}  (0.311, 0.450) &  {\bf 0.235}   (0.158, 0.331)  \\ 
Decay           & $\exp\{\theta_{3kl}\}$ &  {\bf 4.056}   (3.815, 4.300) &  1.092  (0.918, 1.295) &  1.200  (0.921, 1.505) &  0.924   (0.621, 1.327)  \\
\addlinespace
Residual error variance  & $\sigma^{2}_{kl}$      &  0.053   (0.051, 0.054) &  0.050  (0.048, 0.052) &  0.061  (0.057, 0.064) &  0.053   (0.049, 0.057)  \\
\addlinespace
                & $\omega_{11kl}$        &  0.865   (0.806, 0.928) &  0.922  (0.841, 1.011) &  1.025  (0.903, 1.160) &  1.033   (0.866, 1.231)  \\
Covariance      & $\omega_{12kl}$        & {\bf -0.146} (-0.197, -0.094) &  0.022 (-0.083, 0.128) & -0.160 (-0.325, 0.001) & -0.160  (-0.476, 0.150)  \\
matrix for      & $\omega_{13kl}$        &  {\bf 0.595}   (0.521, 0.672) &  {\bf 0.776}  (0.621, 0.946) &  {\bf 0.701}  (0.489, 0.933) &  {\bf 0.655}   (0.314, 1.052)  \\
random effects  & $\omega_{22kl}$        &  0.709   (0.641, 0.781) &  1.134  (0.966, 1.328) &  1.182  (0.961, 1.455) &  1.994   (1.401, 2.842)  \\
                & $\omega_{23kl}$        &  0.044  (-0.017, 0.106) &  {\bf 0.755}  (0.549, 0.981) &  {\bf 0.385}  (0.100, 0.708) &  {\bf 1.424}   (0.770, 2.285)  \\
                & $\omega_{33kl}$        &  1.229   (1.114, 1.350) &  2.621  (2.184, 3.116) &  1.977  (1.509, 2.567) &  2.986   (2.048, 4.321)  \\
\hline
\multicolumn{6}{c}{Free light chains ($k=2$)} \\
\hline
Baseline        & $\exp\{\theta_{1kl}\}$ & {\bf 20.748} (19.340, 22.189) &  {\bf 9.381} (8.662, 10.178) & {\bf 10.246} (9.075, 11.575) & {\bf 14.002} (11.646, 17.098)  \\
Growth          & $\exp\{\theta_{2kl}\}$ &  {\bf 0.162}   (0.149, 0.177) &  {\bf 0.342}  (0.300, 0.388) &  {\bf 0.426}  (0.350, 0.518) &  {\bf 0.392}   (0.284, 0.517)  \\
Decay           & $\exp\{\theta_{3kl}\}$ &  {\bf 2.903}   (2.651, 3.164) &  0.916  (0.736, 1.121) &  1.109  (0.797, 1.499) &  0.997   (0.590, 1.589)  \\
\addlinespace
Residual error variance  & $\sigma^{2}_{kl}$      &  0.100   (0.097, 0.102) &  0.075  (0.072, 0.077) &  0.080  (0.076, 0.083) &  0.102   (0.095, 0.109)  \\
\addlinespace
                & $\omega_{11kl}$        &  3.017   (2.841, 3.198) &  2.330  (2.150, 2.524) &  2.525  (2.248, 2.827) &  2.792   (2.375, 3.275)  \\
Covariance      & $\omega_{12kl}$        & {\bf -1.199} (-1.332, -1.072) &  0.076 (-0.083, 0.234) &  0.247 (-0.023, 0.518) & -0.052  (-0.464, 0.358)  \\
matrix for      & $\omega_{13kl}$        &  {\bf 1.914}   (1.749, 2.095) &  {\bf 1.874}  (1.616, 2.148) &  {\bf 1.851}  (1.463, 2.289) &  {\bf 2.099}   (1.453, 2.872)  \\
random effects  & $\omega_{22kl}$        &  1.477   (1.340, 1.623) &  1.928  (1.701, 2.185) &  2.335  (1.945, 2.791) &  2.263   (1.725, 2.941)  \\
                & $\omega_{23kl}$        &  0.038  (-0.105, 0.183) &  {\bf 1.354}  (1.055, 1.683) &  {\bf 1.582}  (1.102, 2.144) &  {\bf 1.025}   (0.293, 1.875)  \\
                & $\omega_{33kl}$        &  2.980   (2.726, 3.268) &  4.473  (3.821, 5.210) &  4.364  (3.415, 5.525) &  5.008   (3.518, 7.075)  \\
\hline
\end{tabular}
\end{table}

\end{landscape}

\pagebreak

\begin{landscape} \thispagestyle{empty}

\begin{table}[htp!]
\caption{Posterior summary (mean and 95\% credible interval) of the bi-exponential submodel parameters for each biomarker by line of therapy (LoT) using the corrected two-stage (TS) approach. Statistically significant variables are shown in bold, except for variance parameters.} \label{table:postbiexpTS}
\centering
\begin{tabular}{cccccc}
\hline
{\bf Interpretation} & {\bf Parameter}  & {\bf LoT $\bm{l=1}$} &  {\bf LoT $\bm{l=2}$} &  {\bf LoT $\bm{l=3}$} &  {\bf LoT $\bm{l=4}$} \\ 
\hline
\multicolumn{6}{c}{M-spike ($k=1$)} \\
\hline  
Baseline        & $\exp\{\theta_{1kl}\}$ & {\bf 16.841} (16.197, 17.546) &  {\bf 7.675}  (7.250, 8.160) &  {\bf 7.388}   (6.792, 8.047) &  {\bf 7.786}  (6.920, 8.759)  \\
Growth          & $\exp\{\theta_{2kl}\}$ &  {\bf 0.196}   (0.185, 0.207) &  {\bf 0.220}  (0.191, 0.249) &  {\bf 0.283}   (0.231, 0.344) &  {\bf 0.215}  (0.139, 0.306)  \\
Decay           & $\exp\{\theta_{3kl}\}$ &  {\bf 3.624}   (3.401, 3.858) &  0.910  (0.763, 1.080) &  1.049   (0.827, 1.310) &  0.874  (0.562, 1.283)  \\
\addlinespace
Residual error variance  & $\sigma^{2}_{kl}$      &  0.053   (0.051, 0.054) &  0.051  (0.049, 0.053) &  0.061   (0.058, 0.064) &  0.053  (0.049, 0.057)  \\
\addlinespace
                & $\omega_{11kl}$        &  0.865   (0.808, 0.927) &  0.917  (0.838, 1.004) &  1.013   (0.894, 1.146) &  1.030  (0.860, 1.232)  \\
Covariance      & $\omega_{12kl}$        & {\bf -0.135} (-0.193, -0.079) & -0.035 (-0.162, 0.087) & {\bf -0.206} (-0.389, -0.024) & -0.165 (-0.485, 0.147)  \\
matrix for      & $\omega_{13kl}$        &  {\bf 0.631}   (0.554, 0.711) &  {\bf 0.766}  (0.604, 0.941) &  {\bf 0.672}   (0.452, 0.909) &  {\bf 0.663}  (0.316, 1.063)  \\
random effects  & $\omega_{22kl}$        &  0.661   (0.592, 0.737) &  1.292  (1.079, 1.537) &  1.358   (1.066, 1.705) &  2.031  (1.407, 2.931)  \\
                & $\omega_{23kl}$        &  {\bf 0.196}   (0.124, 0.272) &  {\bf 0.981}  (0.735, 1.258) &  {\bf 0.595}   (0.265, 0.977) &  {\bf 1.456}  (0.763, 2.399)  \\
                & $\omega_{33kl}$        &  1.422   (1.294, 1.562) &  2.911  (2.449, 3.449) &  2.173   (1.657, 2.793) &  3.028  (2.018, 4.394)  \\
\hline
\multicolumn{6}{c}{Free light chains ($k=2$)} \\
\hline
Baseline        & $\exp\{\theta_{1kl}\}$ & {\bf 19.802} (18.528, 21.254) &  {\bf 8.915}  (8.184, 9.663) &  {\bf 9.976}  (8.779, 11.311) & {\bf 13.173} (10.983, 15.915)  \\
Growth          & $\exp\{\theta_{2kl}\}$ &  {\bf 0.143}   (0.132, 0.155) &  {\bf 0.178}  (0.148, 0.210) &  {\bf 0.196}   (0.142, 0.260) &  {\bf 0.283}   (0.198, 0.381)  \\
Decay           & $\exp\{\theta_{3kl}\}$ &  {\bf 2.605}   (2.374, 2.850) &  {\bf 0.506}  (0.400, 0.631) &  {\bf 0.642}   (0.445, 0.885) &  0.767   (0.431, 1.222)  \\
\addlinespace
Residual error variance  & $\sigma^{2}_{kl}$      &  0.100   (0.098, 0.102) &  0.075  (0.073, 0.078) &  0.080   (0.077, 0.084) &  0.102   (0.095, 0.109)  \\
\addlinespace
                & $\omega_{11kl}$        &  2.977   (2.810, 3.154) &  2.318  (2.139, 2.511) &  2.546   (2.277, 2.842) &  2.782   (2.365, 3.259)  \\
Covariance      & $\omega_{12kl}$        & {\bf -1.234} (-1.365, -1.105) & -0.106 (-0.333, 0.113) & -0.028  (-0.461, 0.378) & -0.230  (-0.681, 0.211)  \\
matrix for      & $\omega_{13kl}$        &  {\bf 1.897}   (1.729, 2.076) &  {\bf 1.905}  (1.610, 2.223) &  {\bf 1.906}   (1.446, 2.408) &  {\bf 2.053}   (1.393, 2.826)  \\
random effects  & $\omega_{22kl}$        &  1.586   (1.438, 1.741) &  2.464  (2.099, 2.905) &  3.177   (2.494, 4.031) &  2.432   (1.809, 3.270)  \\
                & $\omega_{23kl}$        &  {\bf 0.144}   (0.003, 0.290) &  {\bf 2.031}  (1.571, 2.575) &  {\bf 2.547}   (1.765, 3.531) &  {\bf 1.189}   (0.395, 2.155)  \\
                & $\omega_{33kl}$        &  3.053   (2.787, 3.334) &  5.569  (4.730, 6.598) &  5.713   (4.383, 7.398) &  5.319   (3.698, 7.546)  \\
\hline
\end{tabular}
\end{table}

\end{landscape}

\pagebreak

\begin{landscape} \thispagestyle{empty}

\begin{table}[htp!]
\caption{Posterior summary (mean and 95\% credible interval) for time-to-death ($v=1$) by line of therapy (LoT) using the joint estimation (JE) approach. Statistically significant variables are shown in bold.} \label{table:postdeathJE}
\centering
\begin{tabular}{cccccc}
\hline
{\bf Variable} & {\bf Category}  & {\bf LoT $\bm{l=1}$} &  {\bf LoT $\bm{l=2}$} &  {\bf LoT $\bm{l=3}$} &  {\bf LoT $\bm{l=4}$} \\ 
\hline  
Sex                         & Female          & -0.159  (-0.323, 0.006) & {\bf -0.339} (-0.579, -0.096) & -0.165  (-0.465, 0.134) & -0.307  (-0.627, 0.018)  \\
\addlinespace
\multirow{2}{*}{Ethnicity}  & Non-Hisp. Black & -0.082  (-0.315, 0.143) &  0.004  (-0.326, 0.326) &  0.017  (-0.406, 0.426) &  0.208  (-0.208, 0.602)  \\ 
                            & Other           & -0.012  (-0.237, 0.200) & -0.233  (-0.553, 0.071) & -0.355  (-0.779, 0.050) & {\bf -0.653} (-1.148, -0.177)  \\
\addlinespace
\multirow{2}{*}{ECOG}       & 1               &  0.174  (-0.093, 0.446) &  0.051  (-0.320, 0.423) & -0.069  (-0.517, 0.379) &  {\bf 0.783}   (0.321, 1.271)  \\
                            & 2$^{+}$         &  {\bf 0.806}   (0.539, 1.086) &  {\bf 0.665}   (0.285, 1.050) &  {\bf 0.610}   (0.120, 1.107) &  {\bf 0.963}   (0.358, 1.566)  \\
\addlinespace
\multirow{2}{*}{ISS}        & Stage II        &  {\bf 0.543}   (0.207, 0.877) &  0.197  (-0.227, 0.619) & -0.059  (-0.570, 0.471) &  {\bf 0.617}   (0.090, 1.162)  \\
                            & Stage III       &  {\bf 0.464}   (0.072, 0.858) &  0.316  (-0.162, 0.797) &  0.224  (-0.385, 0.849) &  0.584  (-0.022, 1.214)  \\
\addlinespace
Age                         & --              &  {\bf 0.557}   (0.448, 0.671) &  {\bf 0.472}   (0.322, 0.624) &  {\bf 0.560}   (0.373, 0.755) &  {\bf 0.171}   (0.019, 0.327)  \\ 
Albumin                     & --              & -0.073  (-0.163, 0.023) &  0.056  (-0.125, 0.265) & -0.167  (-0.347, 0.030) & -0.078  (-0.320, 0.205)  \\
B2M                         & --              &  {\bf 0.244}   (0.082, 0.397) & -0.005  (-0.227, 0.200) & -0.268  (-0.588, 0.029) &  0.292  (-0.043, 0.623)  \\
Creatine                    & --              &  0.036  (-0.064, 0.134) &  0.053  (-0.099, 0.200) &  0.055  (-0.152, 0.257) & -0.081  (-0.314, 0.148)  \\
Hemoglobin                  & --              & {\bf -0.125} (-0.220, -0.029) & -0.074  (-0.205, 0.059) & -0.131  (-0.297, 0.034) &  0.091  (-0.089, 0.279)  \\
LDH                         & --              &  {\bf 0.153}   (0.041, 0.263) &  0.140  (-0.025, 0.296) &  0.212  (-0.013, 0.420) & -0.038  (-0.262, 0.173)  \\
Lymphocyte                  & --              & -0.045  (-0.134, 0.045) & -0.030  (-0.166, 0.104) & -0.017  (-0.170, 0.138) & -0.056  (-0.230, 0.121)  \\
Neutrophil                  & --              &  {\bf 0.169}   (0.061, 0.278) &  {\bf 0.178}   (0.018, 0.332) &  0.019  (-0.166, 0.211) &  0.165  (-0.049, 0.380)  \\
Platelet                    & --              & {\bf -0.091} (-0.162, -0.011) & -0.080  (-0.232, 0.072) & -0.178  (-0.357, 0.008) & {\bf -0.318} (-0.513, -0.125)  \\
IgA                         & --              &  0.015  (-0.125, 0.157) &  {\bf 0.279}   (0.089, 0.469) &  0.158  (-0.076, 0.389) & -0.135  (-0.361, 0.093)  \\
IgG                         & --              &  0.129  (-0.010, 0.269) &  0.172  (-0.033, 0.376) & -0.024  (-0.267, 0.221) & {\bf -0.273} (-0.492, -0.050)  \\
IgM                         & --              &  {\bf 0.163}   (0.043, 0.279) &  0.078  (-0.091, 0.227) & -0.076  (-0.324, 0.147) & -0.116  (-0.310, 0.082)  \\
Time prev. LoT              & --              &  --                     & -0.043  (-0.169, 0.078) & {\bf -0.188} (-0.388, -0.004) & {\bf -0.298} (-0.578, -0.039)  \\
\addlinespace
Baseline M-spike ($\alpha_{1lk1}$) & --              &  {\bf 0.206}   (0.055, 0.359) &  {\bf 0.297}   (0.105, 0.490) &  0.134  (-0.127, 0.397) &  {\bf 0.450}   (0.146, 0.778)  \\
Growth M-spike ($\alpha_{1lk2}$) & --              &  {\bf 0.382}   (0.197, 0.566) & -0.049  (-0.293, 0.201) &  0.198  (-0.142, 0.530) &  0.240  (-0.181, 0.680)  \\
Decay M-spike ($\alpha_{1lk3}$) & --              & -0.014  (-0.145, 0.116) & -0.036  (-0.191, 0.117) &  0.053  (-0.201, 0.304) & -0.111  (-0.445, 0.205)  \\
\addlinespace
Baseline FLC ($\alpha_{2lk1}$) & --              &  0.103  (-0.035, 0.236) &  {\bf 0.290}   (0.162, 0.413) &  {\bf 0.253}   (0.077, 0.428) &  {\bf 0.221}   (0.055, 0.390)  \\
Growth FLC ($\alpha_{2lk2}$) & --              &  0.107  (-0.069, 0.273) &  {\bf 0.476}   (0.307, 0.641) &  0.164  (-0.054, 0.383) &  {\bf 0.413}   (0.212, 0.609)  \\
Decay FLC ($\alpha_{2lk3}$) & --              &  0.099  (-0.014, 0.215) & -0.001  (-0.131, 0.135) & -0.070  (-0.263, 0.117) &  0.017  (-0.162, 0.195)  \\
\hline
\end{tabular}
\end{table}

\end{landscape}

\pagebreak

\begin{landscape} \thispagestyle{empty}

\begin{table}[htp!]
\caption{Posterior summary (mean and 95\% credible interval) for time-to-death ($v=1$) by line of therapy (LoT) using the corrected two-stage (TS) approach. Statistically significant variables are shown in bold.} \label{table:postdeathTS}
\centering
\begin{tabular}{cccccc}
\hline
{\bf Variable} & {\bf Category}  & {\bf LoT $\bm{l=1}$} &  {\bf LoT $\bm{l=2}$} &  {\bf LoT $\bm{l=3}$} &  {\bf LoT $\bm{l=4}$} \\ 
\hline  
Sex                         & Female          & -0.155  (-0.316, 0.004) & {\bf -0.315} (-0.547, -0.078) & -0.152  (-0.438, 0.143) & -0.290  (-0.598, 0.017)  \\
\addlinespace
\multirow{2}{*}{Ethnicity}  & Non-Hisp. Black & -0.087  (-0.315, 0.152) & -0.009  (-0.319, 0.314) &  0.041  (-0.382, 0.451) &  0.197  (-0.195, 0.588)  \\ 
                            & Other           &  0.002  (-0.210, 0.214) & -0.235  (-0.548, 0.082) & -0.356  (-0.753, 0.039) & {\bf -0.619} (-1.129, -0.144)  \\
\addlinespace
\multirow{2}{*}{ECOG}       & 1               &  0.158  (-0.105, 0.452) &  0.052  (-0.311, 0.438) & -0.039  (-0.519, 0.413) &  {\bf 0.768}   (0.344, 1.253)  \\
                            & 2$^{+}$         &  {\bf 0.789}   (0.523, 1.069) &  {\bf 0.658}   (0.290, 1.064) &  {\bf 0.647}   (0.134, 1.158) &  {\bf 0.933}   (0.359, 1.519)  \\
\addlinespace
\multirow{2}{*}{ISS}        & Stage II        &  {\bf 0.545}   (0.216, 0.888) &  0.209  (-0.204, 0.625) & -0.034  (-0.532, 0.490) &  {\bf 0.588}   (0.050, 1.139)  \\
                            & Stage III       &  {\bf 0.472}   (0.075, 0.874) &  0.368  (-0.101, 0.847) &  0.238  (-0.381, 0.870) &  0.598  (-0.022, 1.206)  \\
\addlinespace
Age                         & --              &  {\bf 0.555}   (0.443, 0.665) &  {\bf 0.457}   (0.307, 0.605) &  {\bf 0.572}   (0.385, 0.771) &  {\bf 0.163}   (0.005, 0.327)  \\ 
Albumin                     & --              & -0.067  (-0.152, 0.026) &  0.046  (-0.133, 0.245) & -0.165  (-0.347, 0.051) & -0.079  (-0.320, 0.209)  \\
B2M                         & --              &  {\bf 0.239}   (0.075, 0.391) & -0.002  (-0.211, 0.198) & -0.292  (-0.597, 0.003) &  0.284  (-0.036, 0.610)  \\
Creatine                    & --              &  0.041  (-0.058, 0.136) &  0.050  (-0.088, 0.190) &  0.054  (-0.151, 0.252) & -0.100  (-0.323, 0.110)  \\
Hemoglobin                  & --              & {\bf -0.123} (-0.211, -0.027) & -0.075  (-0.218, 0.059) & -0.133  (-0.305, 0.038) &  0.071  (-0.105, 0.254)  \\
LDH                         & --              &  {\bf 0.161}   (0.046, 0.274) &  0.145  (-0.016, 0.292) &  0.217  (-0.007, 0.420) & -0.037  (-0.262, 0.170)  \\
Lymphocyte                  & --              & -0.040  (-0.125, 0.044) & -0.001  (-0.123, 0.131) & -0.019  (-0.159, 0.131) & -0.066  (-0.228, 0.110)  \\
Neutrophil                  & --              &  {\bf 0.168}   (0.067, 0.272) &  {\bf 0.178}   (0.022, 0.333) &  0.027  (-0.155, 0.206) &  0.169  (-0.040, 0.367)  \\
Platelet                    & --              & {\bf -0.093} (-0.161, -0.013) & -0.105  (-0.250, 0.050) & {\bf -0.206} (-0.401, -0.007) & {\bf -0.320} (-0.509, -0.133)  \\
IgA                         & --              &  0.020  (-0.121, 0.165) &  {\bf 0.283}   (0.098, 0.468) &  0.171  (-0.063, 0.405) & -0.126  (-0.344, 0.104)  \\
IgG                         & --              &  0.120  (-0.016, 0.256) &  0.179  (-0.018, 0.374) & -0.011  (-0.261, 0.237) & {\bf -0.267} (-0.486, -0.043)  \\
IgM                         & --              &  {\bf 0.162}   (0.042, 0.276) &  0.088  (-0.070, 0.242) & -0.081  (-0.332, 0.150) & -0.117  (-0.294, 0.066)  \\
Time prev. LoT              & --              &  --                     & -0.027  (-0.143, 0.084) & -0.182  (-0.384, 0.003) & {\bf -0.300} (-0.580, -0.064)  \\
\addlinespace
Baseline M-spike ($\alpha_{1lk1}$) & --              &  {\bf 0.197}   (0.030, 0.350) &  {\bf 0.308}   (0.100, 0.510) &  0.140  (-0.116, 0.403) &  {\bf 0.466}   (0.172, 0.758)  \\
Growth M-spike ($\alpha_{1lk2}$) & --              &  {\bf 0.245}   (0.041, 0.438) & -0.065  (-0.298, 0.198) &  0.123  (-0.180, 0.425) &  0.266  (-0.093, 0.658)  \\
Decay M-spike ($\alpha_{1lk3}$) & --              & -0.046  (-0.177, 0.087) & -0.031  (-0.181, 0.125) &  0.013  (-0.216, 0.251) & -0.127  (-0.440, 0.161)  \\
\addlinespace
Baseline FLC ($\alpha_{2lk1}$) & --              &  0.080  (-0.057, 0.218) &  {\bf 0.309}   (0.176, 0.443) &  {\bf 0.236}   (0.058, 0.419) &  {\bf 0.240}   (0.085, 0.400)  \\
Growth FLC ($\alpha_{2lk2}$) & --              &  0.061  (-0.111, 0.231) &  {\bf 0.282}   (0.111, 0.440) &  0.004  (-0.207, 0.225) &  {\bf 0.330}   (0.152, 0.505)  \\
Decay FLC ($\alpha_{2lk3}$) & --              &  0.098  (-0.021, 0.216) & -0.045  (-0.168, 0.081) & -0.055  (-0.241, 0.123) & -0.028  (-0.185, 0.136)  \\
\hline
\end{tabular}
\end{table}

\end{landscape}

\pagebreak

\begin{landscape} \thispagestyle{empty}

\begin{table}[htp!]
\caption{Posterior summary (mean and 95\% credible interval) for time-to-next-LoT ($v=2$) by line of therapy (LoT) using the joint estimation (JE) approach. Statistically significant variables are shown in bold.} \label{table:postnextJE}
\centering
\begin{tabular}{ccccc}
\hline
{\bf Variable} & {\bf Category}  & {\bf LoT $\bm{l=1}$} &  {\bf LoT $\bm{l=2}$} &  {\bf LoT $\bm{l=3}$} \\ 
\hline
Sex                         & Female          & -0.027  (-0.122, 0.069) &  0.005  (-0.130, 0.140) & -0.185  (-0.379, 0.007)  \\
\addlinespace
\multirow{2}{*}{Ethnicity}  & Non-Hisp. Black & {\bf -0.220} (-0.358, -0.086) & -0.109  (-0.297, 0.076) & -0.151  (-0.422, 0.110)  \\ 
                            & Other           & -0.050  (-0.180, 0.076) & {\bf -0.195} (-0.374, -0.014) & {\bf -0.454} (-0.727, -0.183)  \\
\addlinespace
\multirow{2}{*}{ECOG}       & 1               &  0.125  (-0.010, 0.263) &  0.094  (-0.096, 0.286) &  0.058  (-0.219, 0.341)  \\
                            & 2$^{+}$         &  0.161  (-0.008, 0.327) & -0.094  (-0.336, 0.145) &  0.125  (-0.236, 0.487)  \\
\addlinespace
\multirow{2}{*}{ISS}        & Stage II        & -0.027  (-0.179, 0.126) &  0.048  (-0.163, 0.260) & -0.125  (-0.437, 0.189)  \\
                            & Stage III       &  0.090  (-0.102, 0.285) &  0.048  (-0.203, 0.302) &  0.175  (-0.190, 0.534)  \\
\addlinespace
Age                         & --              & {\bf -0.210} (-0.256, -0.164) & {\bf -0.124} (-0.187, -0.063) & {\bf -0.168} (-0.259, -0.076)  \\ 
Albumin                     & --              &  0.005  (-0.054, 0.067) & -0.050  (-0.146, 0.053) & -0.034  (-0.188, 0.132)  \\
B2M                         & --              &  0.068  (-0.035, 0.168) & -0.040  (-0.172, 0.089) & -0.098  (-0.278, 0.078)  \\
Creatine                    & --              & -0.028  (-0.096, 0.037) & -0.048  (-0.144, 0.045) & -0.011  (-0.148, 0.122)  \\
Hemoglobin                  & --              & {\bf -0.067} (-0.125, -0.010) & -0.002  (-0.080, 0.074) &  0.020  (-0.095, 0.134)  \\
LDH                         & --              & -0.065  (-0.140, 0.009) & -0.047  (-0.152, 0.058) &  0.019  (-0.137, 0.173)  \\
Lymphocyte                  & --              &  0.035  (-0.021, 0.091) & -0.043  (-0.125, 0.039) & -0.009  (-0.114, 0.096)  \\
Neutrophil                  & --              & -0.054  (-0.118, 0.011) & -0.015  (-0.106, 0.075) & -0.075  (-0.198, 0.051)  \\
Platelet                    & --              & -0.008  (-0.061, 0.047) & -0.047  (-0.138, 0.044) & -0.075  (-0.196, 0.046)  \\
IgA                         & --              & {\bf -0.101} (-0.178, -0.025) &  0.009  (-0.104, 0.116) & -0.043  (-0.185, 0.099)  \\
IgG                         & --              & {\bf -0.107} (-0.184, -0.030) & -0.045  (-0.155, 0.065) & {\bf -0.174} (-0.319, -0.030)  \\
IgM                         & --              & -0.064  (-0.139, 0.010) & -0.039  (-0.141, 0.056) & -0.099  (-0.244, 0.038)  \\
Time prev. LoT              & --              &  --                     & {\bf -0.113} (-0.190, -0.038) & {\bf -0.166} (-0.286, -0.049)  \\
\addlinespace
Baseline M-spike ($\alpha_{1lk1}$) & --              &  {\bf 0.332}   (0.233, 0.430) &  {\bf 0.311}   (0.186, 0.437) &  {\bf 0.480}   (0.299, 0.663)  \\
Growth M-spike ($\alpha_{1lk2}$) & --              &  {\bf 0.971}   (0.855, 1.086) &  {\bf 0.631}   (0.444, 0.814) &  {\bf 0.569}   (0.375, 0.769)  \\
Decay M-spike ($\alpha_{1lk3}$) & --              &  0.036  (-0.049, 0.122) & -0.070  (-0.180, 0.041) & {\bf -0.205} (-0.379, -0.035)  \\
\addlinespace
Baseline FLC ($\alpha_{2lk1}$) & --              &  {\bf 0.277}   (0.199, 0.354) &  {\bf 0.206}   (0.122, 0.290) &  {\bf 0.256}   (0.139, 0.376)  \\
Growth FLC ($\alpha_{2lk2}$) & --              &  {\bf 0.295}   (0.183, 0.405) &  {\bf 0.483}   (0.369, 0.599) &  {\bf 0.611}   (0.468, 0.759)  \\
Decay FLC ($\alpha_{2lk3}$) & --              & -0.028  (-0.096, 0.040) &  0.080  (-0.012, 0.173) &  0.041  (-0.095, 0.179)  \\
\hline
\end{tabular}
\end{table}

\end{landscape}

\pagebreak

\begin{landscape} \thispagestyle{empty}

\begin{table}[htp!]
\caption{Posterior summary (mean and 95\% credible interval) for time-to-next-LoT ($v=2$) by line of therapy (LoT) using the corrected two-stage (TS) approach. Statistically significant variables are shown in bold.} \label{table:postnextTS}
\centering
\begin{tabular}{ccccc}
\hline
{\bf Variable} & {\bf Category}  & {\bf LoT $\bm{l=1}$} &  {\bf LoT $\bm{l=2}$} &  {\bf LoT $\bm{l=3}$} \\ 
\hline
Sex                         & Female          & -0.027  (-0.119, 0.063) &  0.015  (-0.113, 0.140) & {\bf -0.195} (-0.381, -0.004)  \\
\addlinespace
\multirow{2}{*}{Ethnicity}  & Non-Hisp. Black & {\bf -0.220} (-0.361, -0.092) & -0.127  (-0.327, 0.054) & -0.119  (-0.386, 0.139)  \\ 
                            & Other           & -0.021  (-0.152, 0.094) & {\bf -0.211} (-0.386, -0.042) & {\bf -0.494} (-0.767, -0.222)  \\
\addlinespace
\multirow{2}{*}{ECOG}       & 1               &  0.117  (-0.026, 0.256) &  0.107  (-0.073, 0.284) &  0.076  (-0.203, 0.351)  \\
                            & 2$^{+}$         &  0.152  (-0.022, 0.321) & -0.079  (-0.313, 0.144) &  0.119  (-0.240, 0.469)  \\
\addlinespace
\multirow{2}{*}{ISS}        & Stage II        & -0.002  (-0.152, 0.153) &  0.051  (-0.159, 0.245) & -0.056  (-0.333, 0.244)  \\
                            & Stage III       &  0.101  (-0.079, 0.288) &  0.094  (-0.148, 0.335) &  0.210  (-0.149, 0.568)  \\
\addlinespace
Age                         & --              & {\bf -0.220} (-0.262, -0.178) & {\bf -0.141} (-0.200, -0.080) & {\bf -0.171} (-0.254, -0.087)  \\ 
Albumin                     & --              &  0.009  (-0.047, 0.067) & -0.057  (-0.147, 0.038) & -0.055  (-0.206, 0.101)  \\
B2M                         & --              &  0.062  (-0.033, 0.154) & -0.035  (-0.165, 0.082) & -0.123  (-0.305, 0.043)  \\
Creatine                    & --              & -0.019  (-0.083, 0.045) & -0.051  (-0.145, 0.041) & -0.003  (-0.140, 0.122)  \\
Hemoglobin                  & --              & -0.056  (-0.111, 0.003) & -0.001  (-0.073, 0.070) &  0.019  (-0.091, 0.125)  \\
LDH                         & --              & -0.044  (-0.113, 0.026) & -0.053  (-0.157, 0.042) &  0.023  (-0.118, 0.164)  \\
Lymphocyte                  & --              &  0.035  (-0.021, 0.086) & -0.025  (-0.104, 0.052) & -0.013  (-0.120, 0.091)  \\
Neutrophil                  & --              & -0.050  (-0.111, 0.011) & -0.010  (-0.100, 0.080) & -0.061  (-0.183, 0.066)  \\
Platelet                    & --              & -0.012  (-0.061, 0.038) & -0.067  (-0.157, 0.021) & -0.099  (-0.220, 0.021)  \\
IgA                         & --              & {\bf -0.098} (-0.178, -0.018) &  0.009  (-0.098, 0.115) & -0.027  (-0.170, 0.117)  \\
IgG                         & --              & {\bf -0.122} (-0.202, -0.045) & -0.044  (-0.149, 0.058) & {\bf -0.168} (-0.321, -0.013)  \\
IgM                         & --              & -0.063  (-0.138, 0.013) & -0.031  (-0.132, 0.067) & -0.099  (-0.237, 0.033)  \\
Time prev. LoT              & --              &  --                     & {\bf -0.107} (-0.185, -0.029) & {\bf -0.164} (-0.295, -0.041)  \\
\addlinespace
Baseline M-spike ($\alpha_{1lk1}$) & --              &  {\bf 0.376}   (0.288, 0.473) &  {\bf 0.367}   (0.252, 0.488) &  {\bf 0.507}   (0.340, 0.681)  \\
Growth M-spike ($\alpha_{1lk2}$) & --              &  {\bf 0.803}   (0.692, 0.918) &  {\bf 0.561}   (0.404, 0.724) &  {\bf 0.520}   (0.338, 0.721)  \\
Decay M-spike ($\alpha_{1lk3}$) & --              & -0.063  (-0.137, 0.009) & -0.084  (-0.182, 0.016) & {\bf -0.229} (-0.386, -0.084)  \\
\addlinespace
Baseline FLC ($\alpha_{2lk1}$) & --              &  {\bf 0.269}   (0.192, 0.347) &  {\bf 0.253}   (0.173, 0.337) &  {\bf 0.357}   (0.243, 0.475)  \\
Growth FLC ($\alpha_{2lk2}$) & --              &  {\bf 0.243}   (0.148, 0.337) &  {\bf 0.320}   (0.220, 0.417) &  {\bf 0.452}   (0.313, 0.594)  \\
Decay FLC ($\alpha_{2lk3}$) & --              & -0.037  (-0.105, 0.029) &  0.005  (-0.077, 0.084) & -0.069  (-0.185, 0.047)  \\
\hline
\end{tabular}
\end{table}

\end{landscape}

\pagebreak

\begin{table}[htp!]
\caption{Baseline variables of patients used to illustrate individual dynamic predictions.}
\label{table:dynpredpatients}
\centering
\begin{tabular}{ccc}
\hline
{\bf Baseline variable}               & {\bf Patient A}    & {\bf Patient B} \\
\hline
Sex                                   & Male               & Female             \\
Ethnicity                             & Non-Hispanic white & Non-Hispanic white \\
ECOG                                  & 2$^{+}$            & 0                  \\
ISS                                   & Not reported       & Stage III          \\
Age (years)                           & 72                 & 73                 \\
Albumin (serum, g/L)                  & 34                 & 19                 \\
B2B (serum, mg/L)                     & NA                 & 13.8               \\
Creatinine (serum, mg/dL)             & 0.93               & 0.9                \\
Hemoglobin (g/dL)                     & 13.8               & 9.9                \\
LDH (serum, U/L)                      & Not reported       & 308                \\
Lymphocyte (count, $\times 10^{9}$/L) & 0.847              & 1.5                \\
Neutrophil (count, $\times 10^{9}$/L) & 4.7                & 5.4                \\
Platelet (count, $\times 10^{9}$/L)   & 233                & 215                \\
IgA (serum, g/L)                      & NA                 & 0.18               \\
IgG (serum, g/L)                      & 83                 & 93.68              \\
IgM (serum, g/L)                      & NA                 & 0.15               \\
\hline
\end{tabular}
\end{table}

\pagebreak

\section*{Appendix B. Web Figures}

\begin{figure}[htp!]
  \centering
  \includegraphics[width=0.86\textwidth]{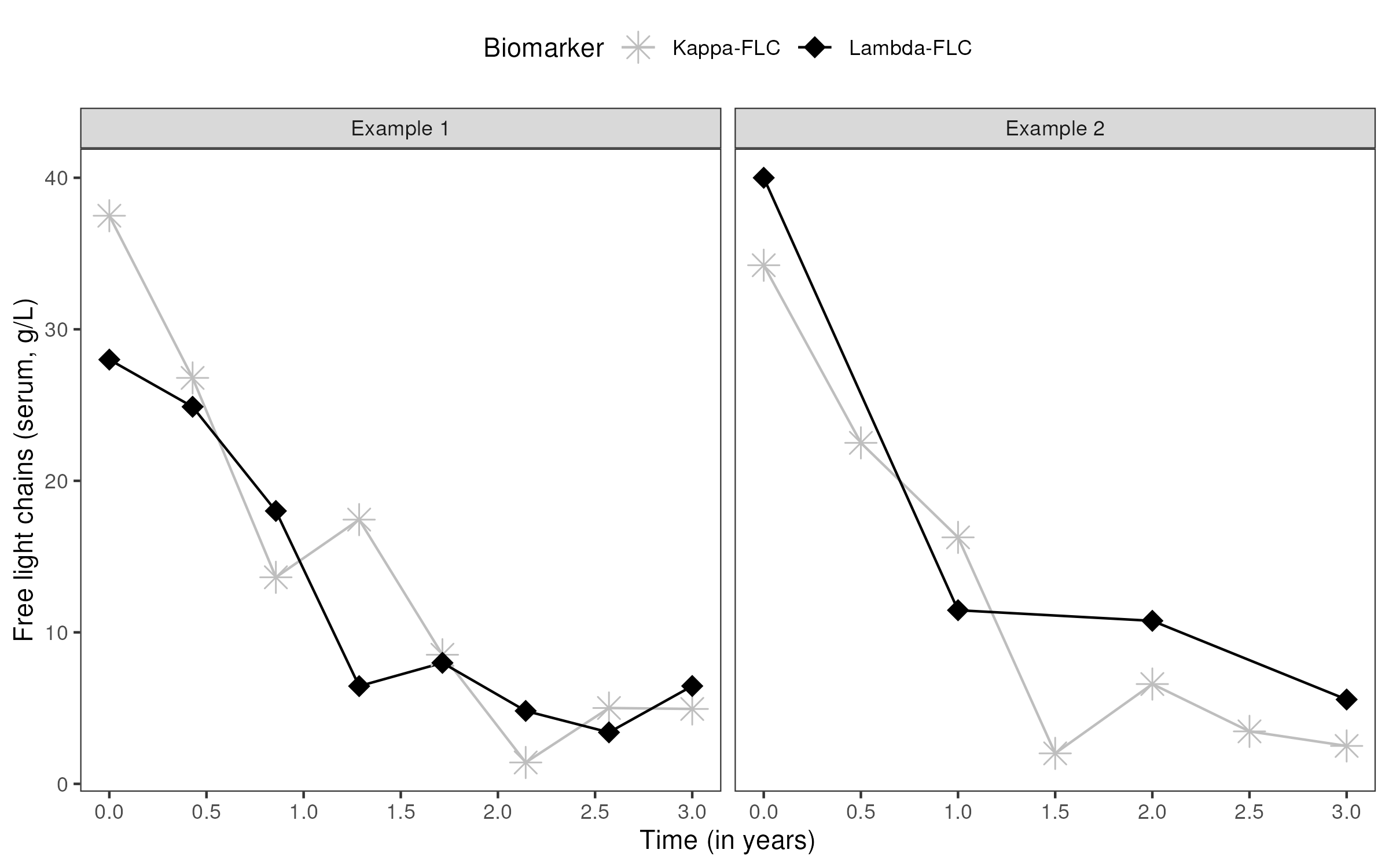}
  \caption{Illustrations of kappa and lambda free light chain (FLC) trajectories. In Example 1 (left), as the first kappa-FLC value is higher than the first lambda-FLC value, only kappa-FLC values are followed through. In Example 2 (right) the opposite occurs, as the first lambda-FLC value is higher than the first kappa-FLC value, only lambda-FLC values are followed through.}
  \label{fig0}
\end{figure}

\begin{figure}[htp!]
  \centering
  \includegraphics[width=0.57\textwidth]{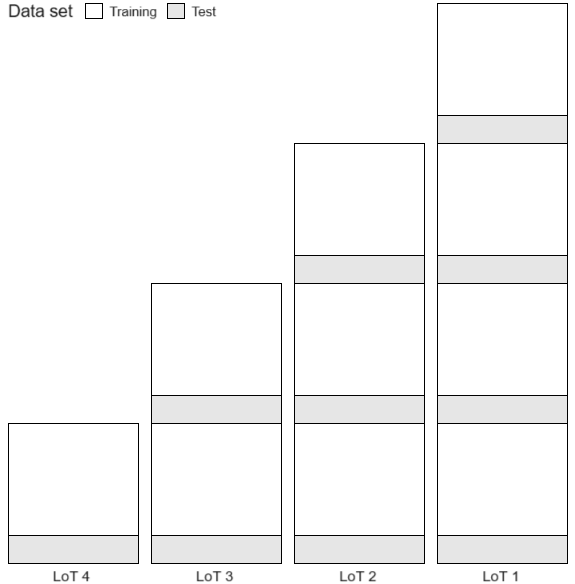}
  \caption{Data splitting scheme into training and test sets by line of therapy (LoT). From bottom to top, first row represents patients in LoTs 1-4, and last one represents patients in LoT 1 but not in LoTs 2-4.}
  \label{fig01}
\end{figure}

\pagebreak

\begin{landscape} \thispagestyle{empty}

\begin{figure}[htp!] \vspace{-1.2cm}
  \centering
  \includegraphics[width=1.4\textwidth]{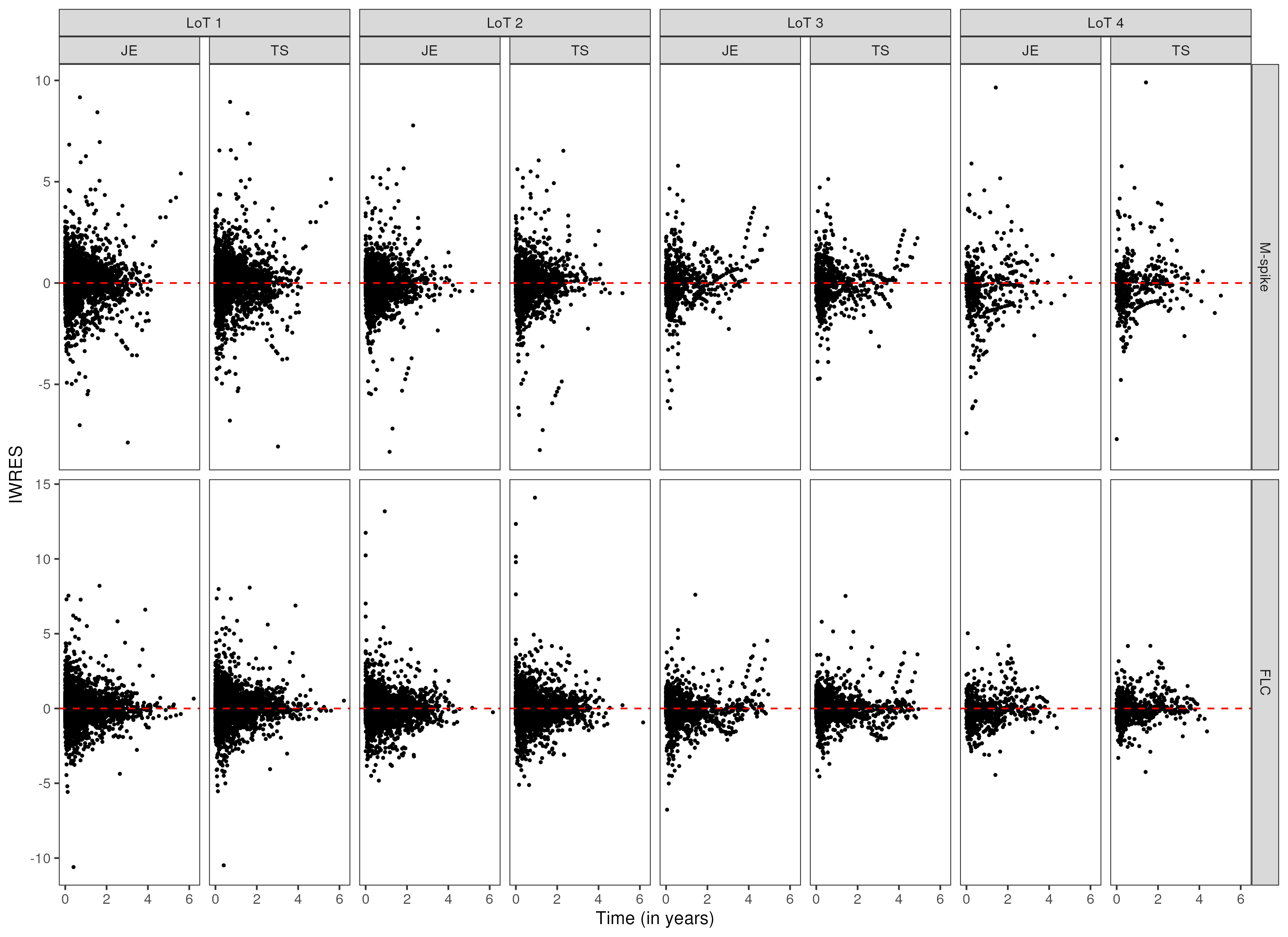}
  \caption{Individual weighted residuals (IWRES) from the bi-exponential submodel for each biomarker considering the test data set by line of therapy (LoT) using joint estimation (JE) and corrected two-stage (TS) approaches. For suitable fit, IWRES should approximately follow a Normal distribution centred at zero \citep{kerioui2022}.}
  \label{fig2}
\end{figure}

\end{landscape}

\pagebreak

\begin{landscape} \thispagestyle{empty}

\begin{figure}[htp!] \vspace{-1.2cm}
  \centering
  \includegraphics[width=1.4\textwidth]{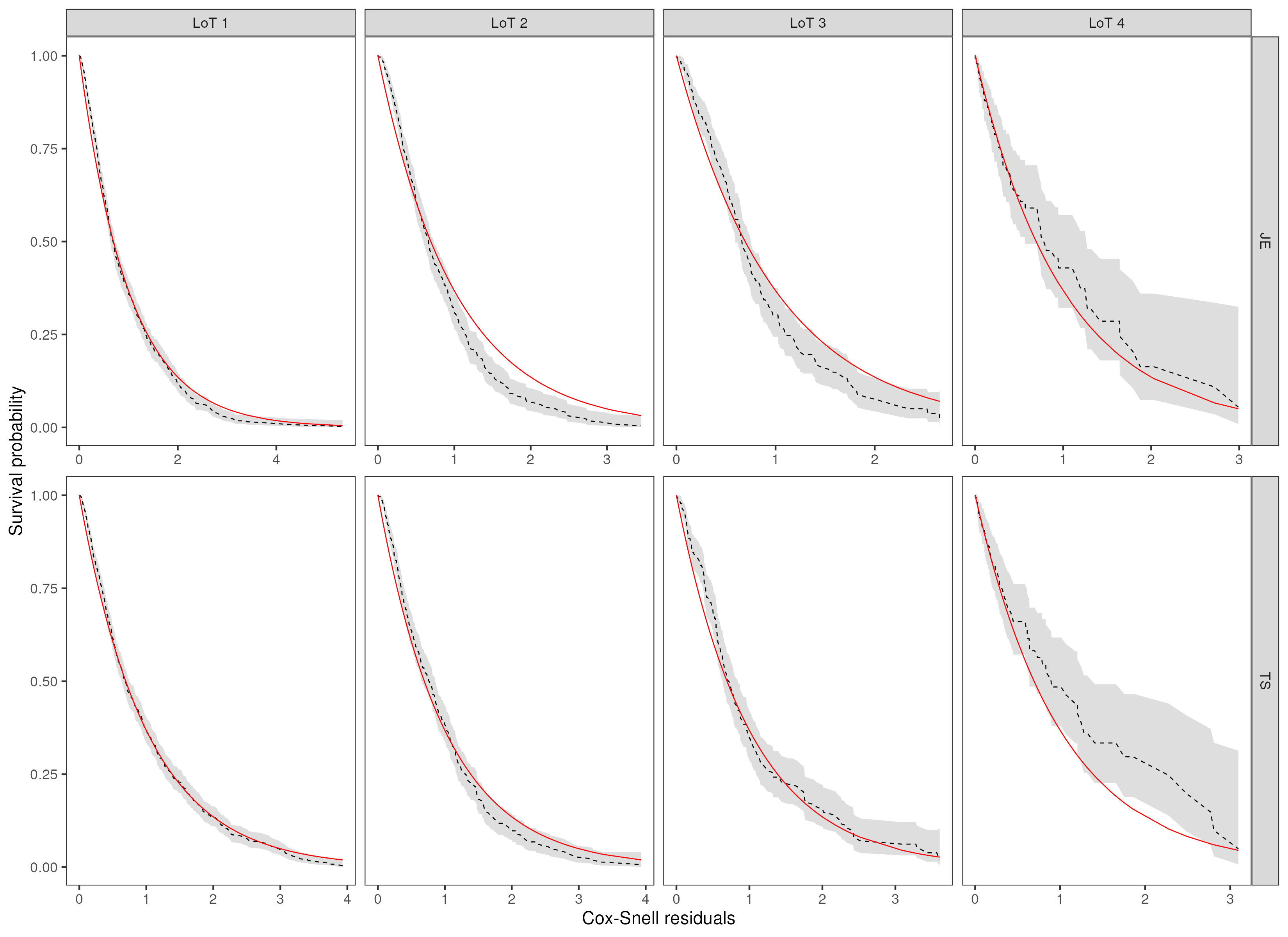}
  \caption{Kaplan–Meier estimates of the Cox–Snell residuals (dashed black line) and its 95\% confidence interval (gray shadow) from the survival submodel considering the test data set by line of therapy (LoT) using joint estimation (JE) and corrected two-stage (TS) approaches. Solid red line: the survival function of the unit exponential distribution.}
  \label{fig3}
\end{figure}

\end{landscape}

\pagebreak

\begin{landscape} \thispagestyle{empty}

\begin{figure}[htp!] \vspace{-1.2cm}
  \centering
  \includegraphics[width=1.4\textwidth]{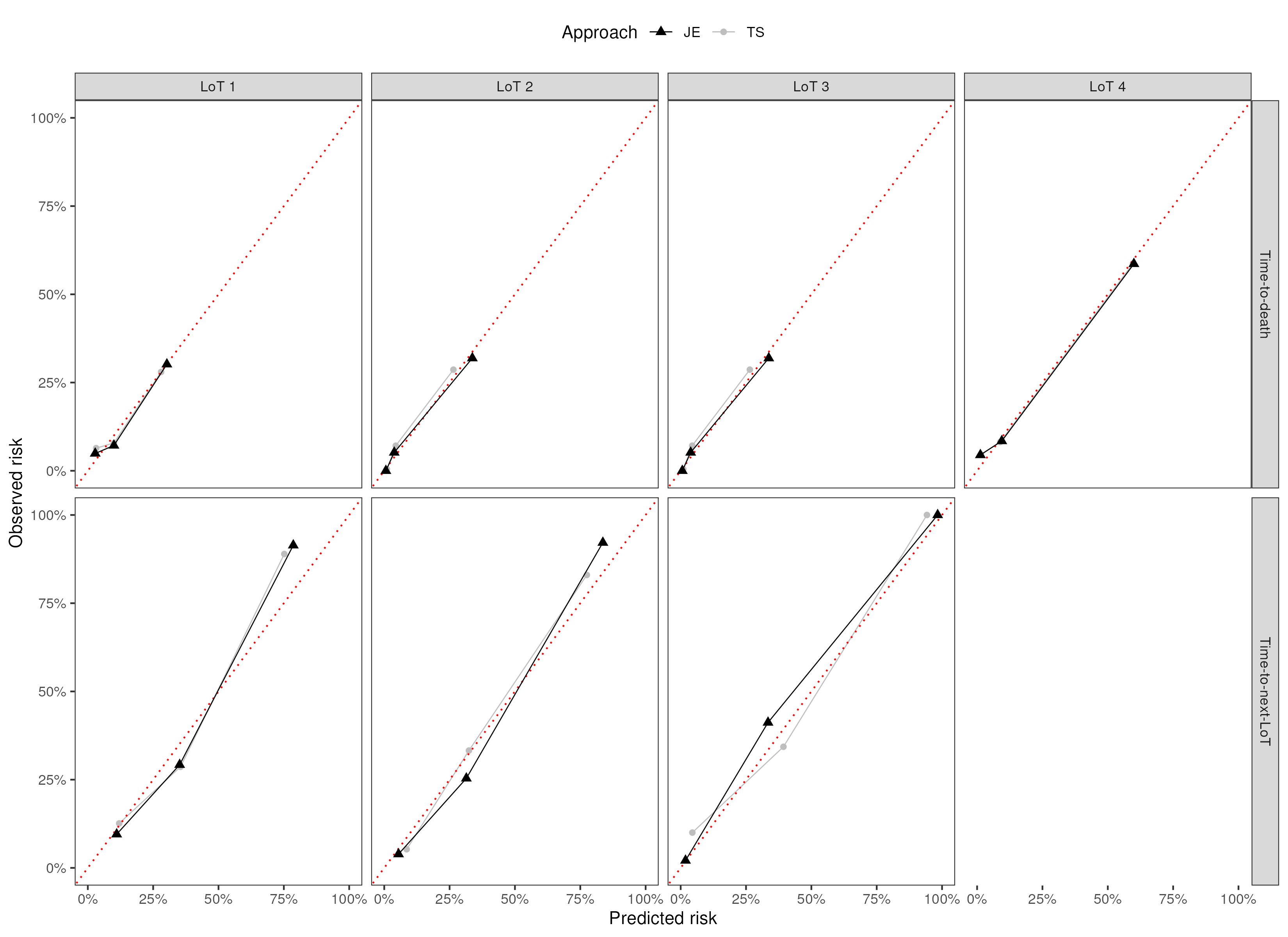}
  \caption{Calibration plots by tercile of predicted 1-year risk from the survival submodel with 6 months of landmark time for each time-to-event considering the test data set by line of therapy (LoT) using joint estimation (JE, black triangle) and corrected two-stage (TS, gray circle) approaches.}
  \label{fig5b}
\end{figure}

\end{landscape}

\pagebreak

\begin{landscape} \thispagestyle{empty}

\begin{figure}[htp!] \vspace{-1.2cm}
  \centering
  \includegraphics[width=1.4\textwidth]{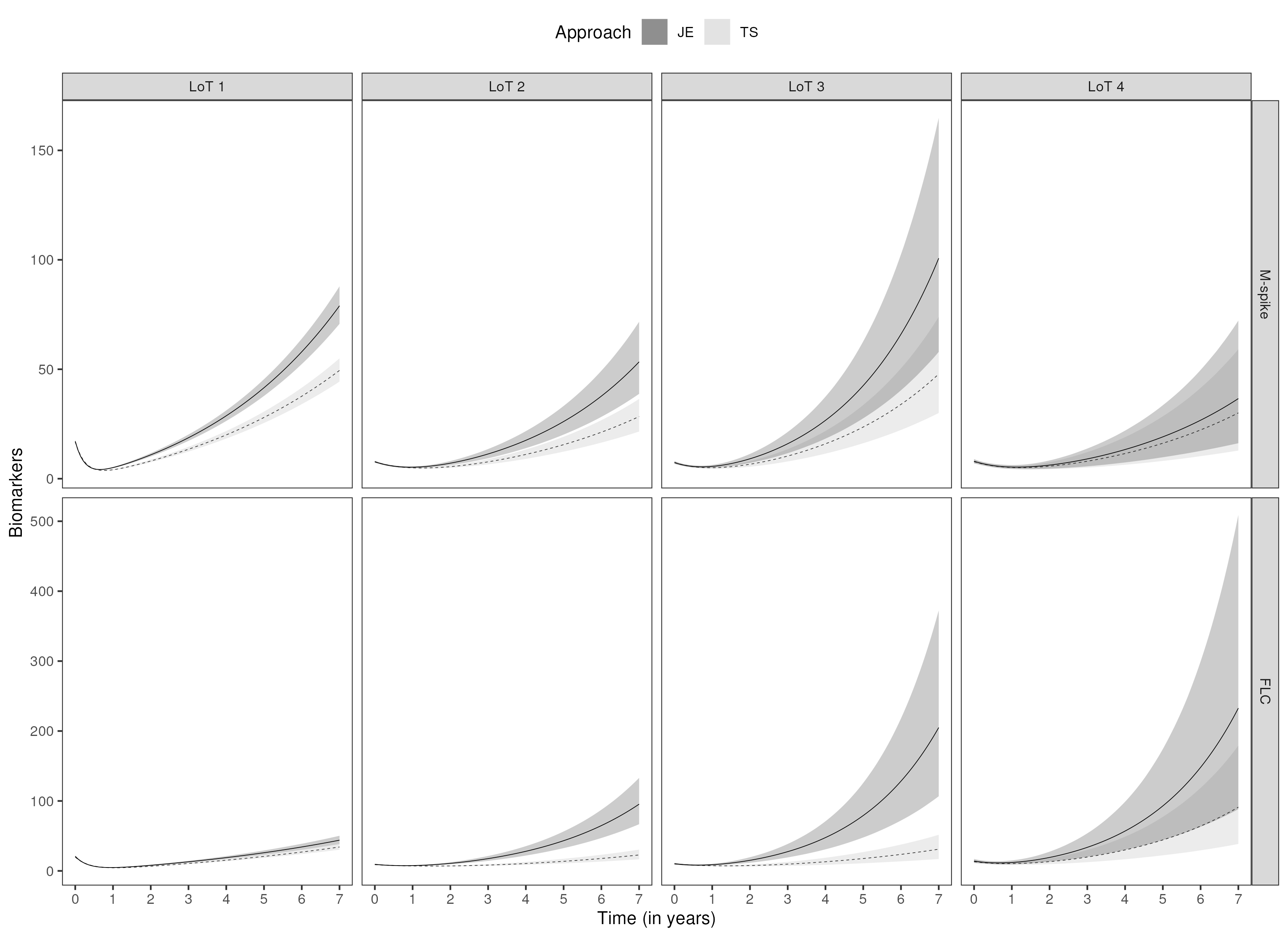}
  \caption{Posterior mean trajectory and its 95\% credible interval from the bi-exponential submodel for each biomarker by line of therapy (LoT) using joint estimation (JE, solid line) and corrected two-stage (TS, dashed line) approaches.}
  \label{fig1}
\end{figure}

\end{landscape}

\pagebreak

\begin{figure}[htp!] \vspace{-1cm}
  \centering
  \includegraphics[width=0.9\textwidth]{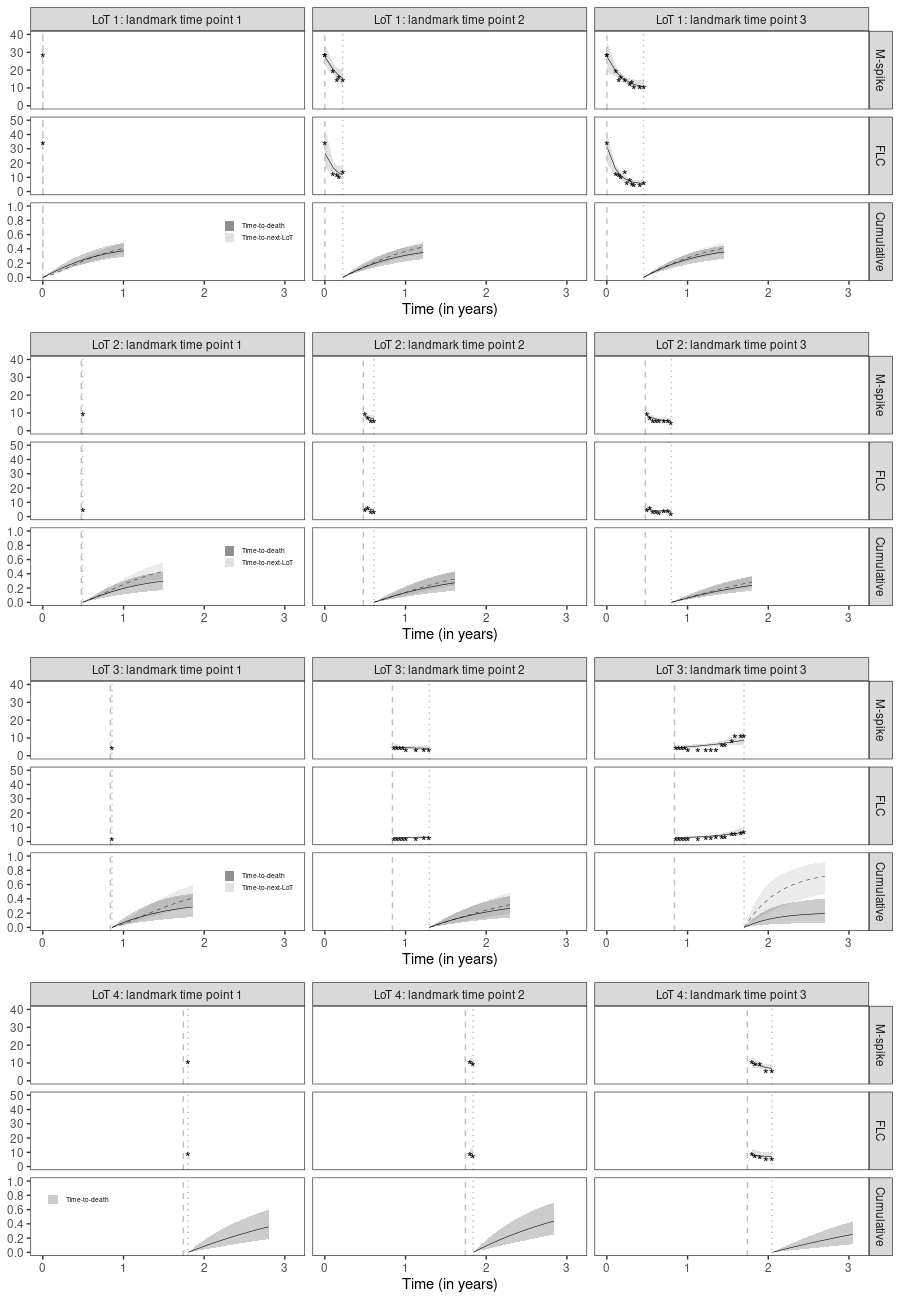}
  \caption{One-year dynamic predictions from three landmark time points (vertical dotted lines) in each line of therapy (LoT) using the joint estimation (JE) approach for patient A. Vertical dashed lines indicate LoT initiation times. For ``M-spike'' and ``FLC'' rows, stars represent biomarker observed values with their respective posterior mean trajectory (solid lines) and 95\% credible intervals (gray shadow). For ``Cumulative'' rows, solid and dashed lines are posterior means (with their respective 95\% credible intervals) of cumulative incidence functions (LoTs 1, 2 \& 3) or distribution functions (LoT 4) for time-to-death and time-to-next-LoT, respectively.}
  \label{fig6JE}
\end{figure}

\pagebreak

\begin{figure}[htp!] \vspace{-1cm}
  \centering
  \includegraphics[width=0.9\textwidth]{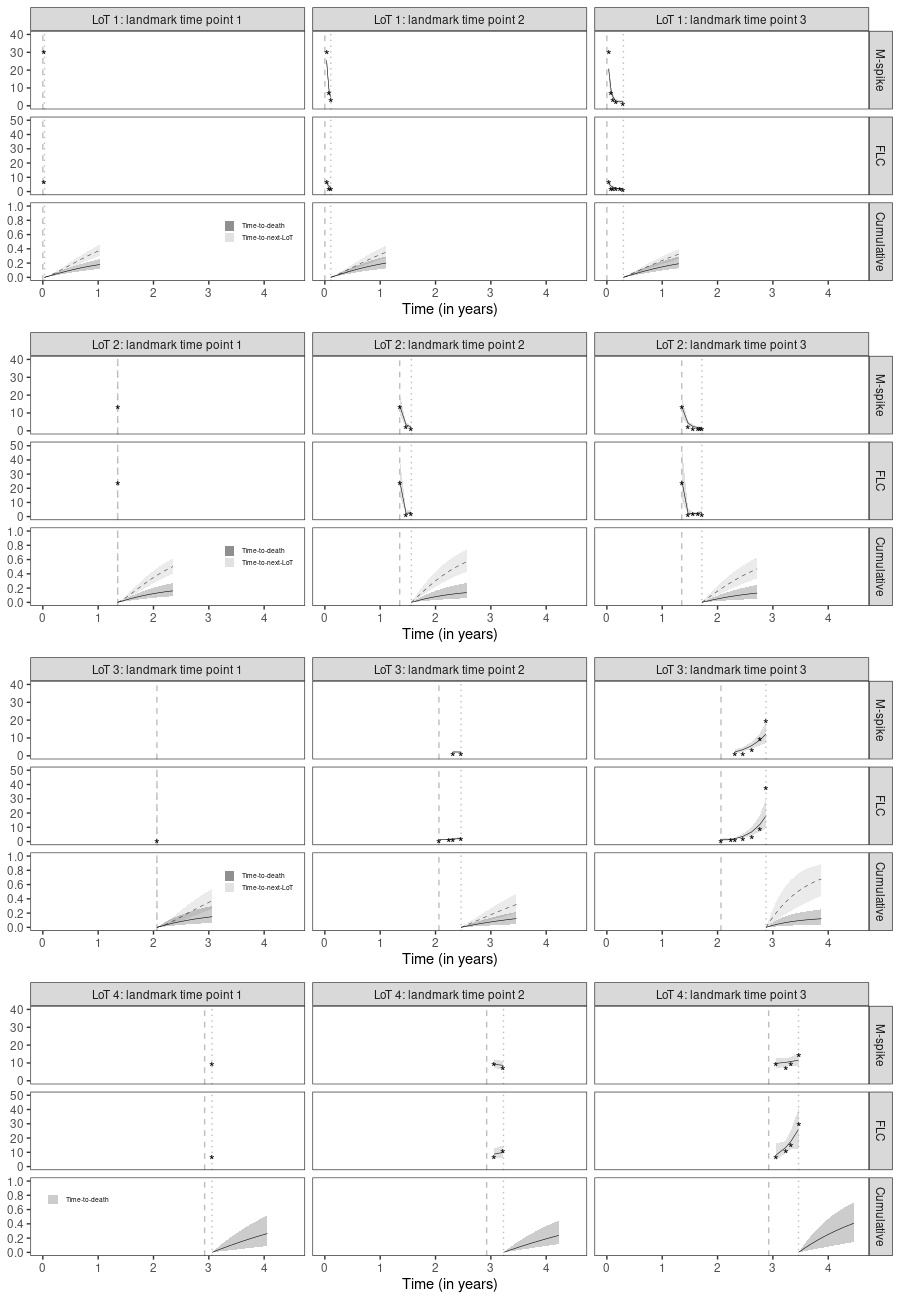}
  \caption{One-year dynamic predictions from three landmark time points (vertical dotted lines) in each line of therapy (LoT) using the joint estimation (JE) approach for patient B. Vertical dashed lines indicate LoT initiation times. For ``M-spike'' and ``FLC'' rows, stars represent biomarker observed values with their respective posterior mean trajectory (solid lines) and 95\% credible intervals (gray shadow). For ``Cumulative'' rows, solid and dashed lines are posterior means (with their respective 95\% credible intervals) of cumulative incidence functions (LoTs 1, 2 \& 3) or distribution functions (LoT 4) for time-to-death and time-to-next-LoT, respectively.}
  \label{fig7JE}
\end{figure}

\end{document}